\documentclass[iop,revtex4, apjfonts]{emulateapj}

\usepackage{amssymb,amsmath,mathrsfs,graphicx}
\usepackage{verbatim}
\usepackage{subfigure}
\usepackage{xcolor}
\usepackage{hyperref}
\tabletypesize{\scriptsize}

\def\h2{$\rm H_2$}

\newcommand{\msun}{M$_{\odot}$}
\newcommand{\lsun}{L$_{\odot}$}

\newcommand{\halpha}{H$\alpha$}

\newcommand{\hi}{H{\sc I}}

\begin{document}

\shortauthors{Weisz et al.}
\title{The Star Formation Histories of Local Group Dwarf Galaxies \sc{I}. Hubble Space Telescope / Wide Field Planetary Camera 2 Observations\altaffilmark{*}}

\author{
Daniel R.\ Weisz\altaffilmark{1,2,7},
Andrew E.\ Dolphin\altaffilmark{3}, 
Evan D.\ Skillman\altaffilmark{4},
Jon Holtzman\altaffilmark{5},
Karoline M. Gilbert\altaffilmark{2,6},
Julianne J.\ Dalcanton\altaffilmark{2},
Benjamin F. Williams\altaffilmark{2}
}

\altaffiltext{*}{Based on observations made with the NASA/ESA Hubble Space Telescope, obtained from the Data Archive at the Space Telescope Science Institute, which is operated by the Association of Universities for Research in Astronomy, Inc., under NASA constract NAS 5-26555.}
\altaffiltext{1}{Department of Astronomy, University of California at Santa Cruz, 1156 High Street, Santa Cruz, CA, 95064 USA; drw@ucsc.edu}
\altaffiltext{2}{Department of Astronomy, University of Washington, Box 351580, Seattle, WA 98195, USA}
\altaffiltext{3}{Raytheon Company, 1151 East Hermans Road, Tucson, AZ 85756, USA}
\altaffiltext{4}{Minnesota Institute for Astrophysics, University of Minnesota, 116 Church Street SE, Minneapolis, MN 55455, USA}
\altaffiltext{5}{Department of Astronomy, New Mexico State University, Box 30001, 1320 Frenger St., Las Cruces, NM 88003}
\altaffiltext{6}{Space Telescope Science Institute, 3700 San Martin Drive, Baltimore, MD, 21218, USA}
\altaffiltext{7}{Hubble Fellow}

\begin{abstract}

We present uniformly measured star formation histories (SFHs) of 40 Local Group dwarf galaxies based on color-magnitude diagram (CMD) analysis from archival Hubble Space Telescope imaging. We demonstrate that accurate SFHs can be recovered from CMDs that do not reach the oldest main sequence turn-off (MSTO), but emphasize that the oldest MSTO is critical for precisely constraining the earliest epochs of star formation.  We find that:  (1) the average lifetime SFHs of dwarf spheroidals (dSphs) can be approximated by an exponentially declining SFH with $\tau$ $\sim$ 5 Gyr; (2) lower luminosity dSphs are less likely to have extended SFHs than more luminous dSphs; (3) the average SFHs of dwarf irregulars (dIrrs), transition dwarfs (dTrans), and dwarf ellipticals (dEs) can be approximated by the combination of an exponentially declining SFH ($\tau$ $\sim$ 3-4 Gyr) for lookback ages $>$ 10-12 Gyr ago and a constant SFH thereafter;  (4) the observed fraction of stellar mass formed prior to z=2 ranges considerably (80\% for galaxies with M $<$ 10$^5$ M$_{\odot}$ to 30\% for galaxies with M$>$10$^7$ M$_{\odot}$) and is largely explained by environment; (5) the distinction between "ultra-faint" and "classical" dSphs is arbitrary; (6) LG dIrrs formed a significantly higher fraction of stellar mass prior to z=2 than the SDSS galaxies from Leiter 2012 and the SFHs from the abundance matching models of Behroozi et al. 2013.  This may indicate higher than expected star-formation efficiencies at early times in low mass galaxies. Finally, we provide all the SFHs in tabulated electronic format for use by the community.

\end{abstract}


\section{Introduction}
\label{intro}

The Local Group (LG) is home to a diverse collection of dwarf galaxies.  They span a wide range in luminosity, metallicity, and stellar content, all of which encode information about the origin and evolutionary history of each galaxy, and in turn, the assembly history of the LG as a whole \citep[e.g.,][]{hodge1971, hodge1989, mateo1998, vandenberg2000, tolstoy2009, mcconnachie2012, ibata2013}.

Because of their low masses, dwarfs are recognized as relatively fragile systems. Present day dwarfs have been strongly influenced by both external processes (e.g., tidal effects, ram pressure, intense ultra-violet radiation from the epoch of cosmic reionization and from star formation in nearby massive galaxies) and internal mechanisms (e.g., star formation, stellar feedback from supernovae, galactic scale winds, turbulence). These processes have re-shaped their dark and baryonic components throughout the galaxies' lifetimes \citep[e.g.,][]{dekel1986, efstathiou1992, maclow1999, bullock2000, mayer2001, governato2010, brooks2013, kazantzidis2013}.  Disentangling the relative contribution of each physical process is key to understanding a wide range of astrophysics including star formation in extremely low metallicity environments, galaxy morphological transformations, the coupling of dark and baryonic matter, the mechanisms by which galaxies' gas supplies are depleted, interactions between host and satellite systems, and the mass assembly history of the LG, all of which have broader implications for how galaxies and groups of galaxies evolve.  

At present, all known dwarf galaxies in the LG have been observed with space- and/or ground-based imaging.  These data reveal the galaxies' resolved stellar populations \citep[e.g.,][and references therein]{hodge1989, mateo1998, dejong2008a, tolstoy2009, mcconnachie2012, brown2012, belokurov2013, martin2013}, allowing us to infer their stellar age distributions. The colors and luminosities of individual stars can be directly compared to stellar evolution models, yielding the galaxy's star formation rate (SFR) as a function of look back time and metallicity, i.e., a star formation history (SFH).  Within the literature, there are approximately two dozen algorithms for this resolved star color-magnitude diagram (CMD) analysis,  Most have been shown to be produce consistent solutions when applied to the same data, providing confidence in our overall ability to accurately decipher CMDs \citep[c.f.,][]{skillman2002, skillman2003, tosi2007, tolstoy2009, monelli2010c, walmswell2013}.    

\begin{figure}
\begin{center}
\plotone{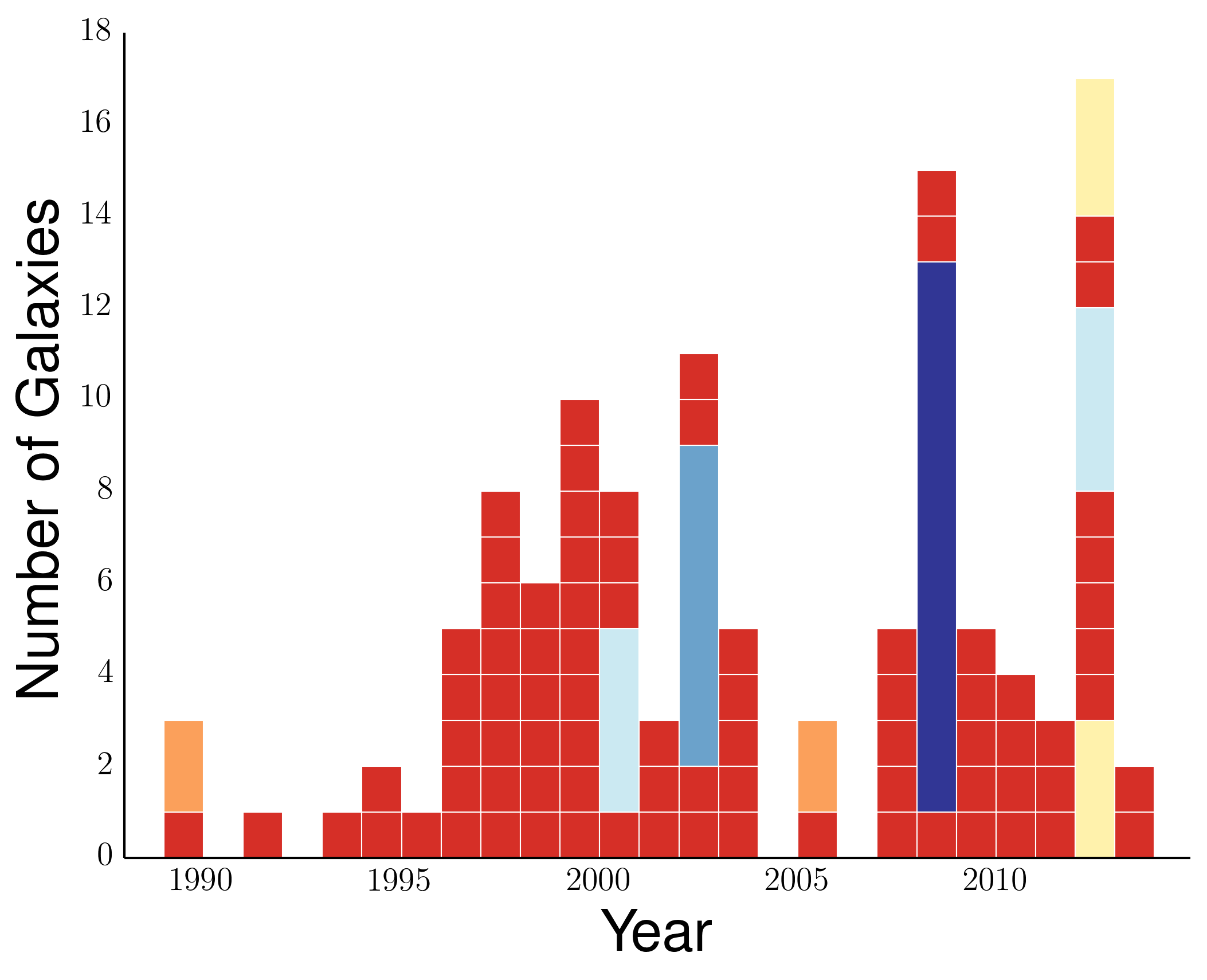}
\caption{A visualization of past published CMD-based SFH studies of LG dwarf galaxies between 1989 and 2013.  Each box represents a single paper and its height indicates the number of galaxies included in the paper.  The colors range from single galaxy papers in dark red to a paper with 12 galaxies in dark blue.  As illustrated here, the majority of past LG dwarf SFHs are from papers studying single or small sub-samples, resulting in a fair degree of heterogeneity in our current knowledge of LG dwarf SFHs.  We have excluded LG SFH compilations \citep[e.g.,][]{mateo1998, tolstoy2009}, as they only provide summaries of SFHs already in the literature. The literature references for this plot are listed in Table \ref{tab:litsfhs}.}
\label{fig:literature_sfhs}
\end{center}
\end{figure}

However, despite extensive observations and analyses, most SFHs have been derived in studies of single or small sub-samples of LG dwarfs, as shown in Figure \ref{fig:literature_sfhs} and Table \ref{tab:litsfhs}.  The majority of these studies use different data reduction methods and CMD analysis techniques ranging from overlying isochrones by eye to formal CMD fitting.  They also employ different stellar evolution libraries, and present results in a wide variety of formats.  The result is a rich but extremely heterogenous measurement of LG dwarf SFHs.  Some of these limitations have been addressed in compilations and review articles \citep[e.g.,][]{grebel1997, mateo1998, tosi2007, tolstoy2009}, but the overall lack of uniformity remains. 

One way to remedy these shortcomings is by uniformly analyzing large datasets of high quality imaging.  Here, we focus on data taken with the Hubble Space Telescope (HST), which has resolved millions of stars in dozens of LG dwarf galaxies over the past 20 years \citep[e.g.,][]{holtzman2006, monelli2010b, monelli2010c, hidalgo2011, brown2012}. The instrumental uniformity and characteristically deep CMDs provided by HST are therefore among the best available for systematically measuring SFHs of LG dwarf galaxies.  HST's high angular resolution allows one to measure fluxes for stars fainter than the current generation of ground-based telescopes, with which we are unable to resolve stars fainter than the oldest main sequence turnoff (MSTO) in any but the Milky Way (MW) satellites.

In this paper, we take a first step toward increasing the uniformity of LG analyses by presenting SFHs for 40 LG dwarfs.  We carry out uniform reduction and analysis of archival observations taken with the HST/Wide Field Planetary Camera 2 \citep[WFPC2;][]{holtzman1995};  Although no longer in service, HST/WFPC2 was used to image a large fraction of the LG dwarf population, providing deep CMDs for nearly all targeted systems.  Presently, we focus on deriving SFHs for LG dwarfs using homogeneously reduced datasets and an identical SFH code, thereby providing an extensive dataset of SFHs with minimal systematics.  This work builds on our preliminary, but unpublished, HST-based LG dwarf SFHs first presented in \citet{dolphin2005}, and later summarized in \citet{orban2008} and \citet{weisz2011b}.  In a subsequent paper in this series, we will include SFHs derived from recent and ongoing observations with the Advanced Camera for Surveys \citep[ACS;][]{ford1998} and Wide Field Camera 3 \citep[WFC3;][]{kimble2008}, which are acquiring imaging of dwarfs without previous HST pointings and/or are increasing the CMD depths in order to fully capture the oldest MSTO for galaxies with existing HST observations \citep[e.g.,][]{cole2007,monelli2010b, monelli2010c,hidalgo2011,brown2012}.  

\begin{deluxetable}{ll}
\tablecolumns{2}
\tablecaption{Literature SFHs of Local Group Dwarf Galaxies}
\tablehead{
\colhead{Reference} &
\colhead{Galaxy} \\
\colhead{(1)} &
\colhead{(2)}
}
\startdata
\citet {ferraro1989} & WLM \\
 \citet{tosi1989} & WLM, Sex~B \\
\citet{tosi1991} & Sex~B \\
\citet{greggio1993} & DDO~210 \\
\citet{smeckerhane1994} & Carina \\
 \citet{marconi1995} & NGC~6822 \\
 \citet{dacosta1996} & And~{\sc I} \\
\citet{gallart1996} & NGC~6822 \\
\citet{lee1996} & NGC~205 \\

\citet{saviane1996} & Tucana \\
 \citet{tolstoy1996} & Leo~A \\
 \citet{aparicio1997a} & LGS3 \\
 \citet{aparicio1997b} & Peg~DIG \\
 \citet{aparicio1997c} & Antlia \\
\citet{dohmpalmer1997a} & Sex~A \\
\citet{dohmpalmer1997b} & Sex~A \\

\citet {han1997} & NGC~147 \\
 \citet{mighell1997} & Carina \\
\citet{mould1997} & LGS3 \\
\citet{dohmpalmer1998} & GR~8 \\
 \citet{gallagher1998} & Peg~DIG \\
 \citet{grillmair1998} & Draco \\
\citet{hurleykeller1998} & Carina \\
\citet{martinezdelgado1998} & NGC~185 \\

\citet {tolstoy1998} & Leo~A \\
 \citet{bellazzini1999} & Sag~DIG \\
\citet{buonanno1999} & Fornax \\
\citet{caputo1999} & Leo {\sc I} \\
 \citet{cole1999} & IC~1613 \\
 \citet{gallart1999} & Leo {\sc I} \\
\citet{karachentsev1999} & Sag~DIG \\
\citet{martinezdelgado1999} & NGC~185

\enddata
\tablecomments{An exhaustive list of original CMD-based SFH studies of LG dwarfs between 1989 and 2013.  We have excluded LG SFH compilations \citep[e.g.,][]{mateo1998, tolstoy2009}, as they only provide summaries of SFHs already in the literature. The full list is shown in Table \ref{tab:full_litsfhs}.}
\label{tab:litsfhs}
\end{deluxetable}

\begin{figure*}
\begin{center}
\plotone{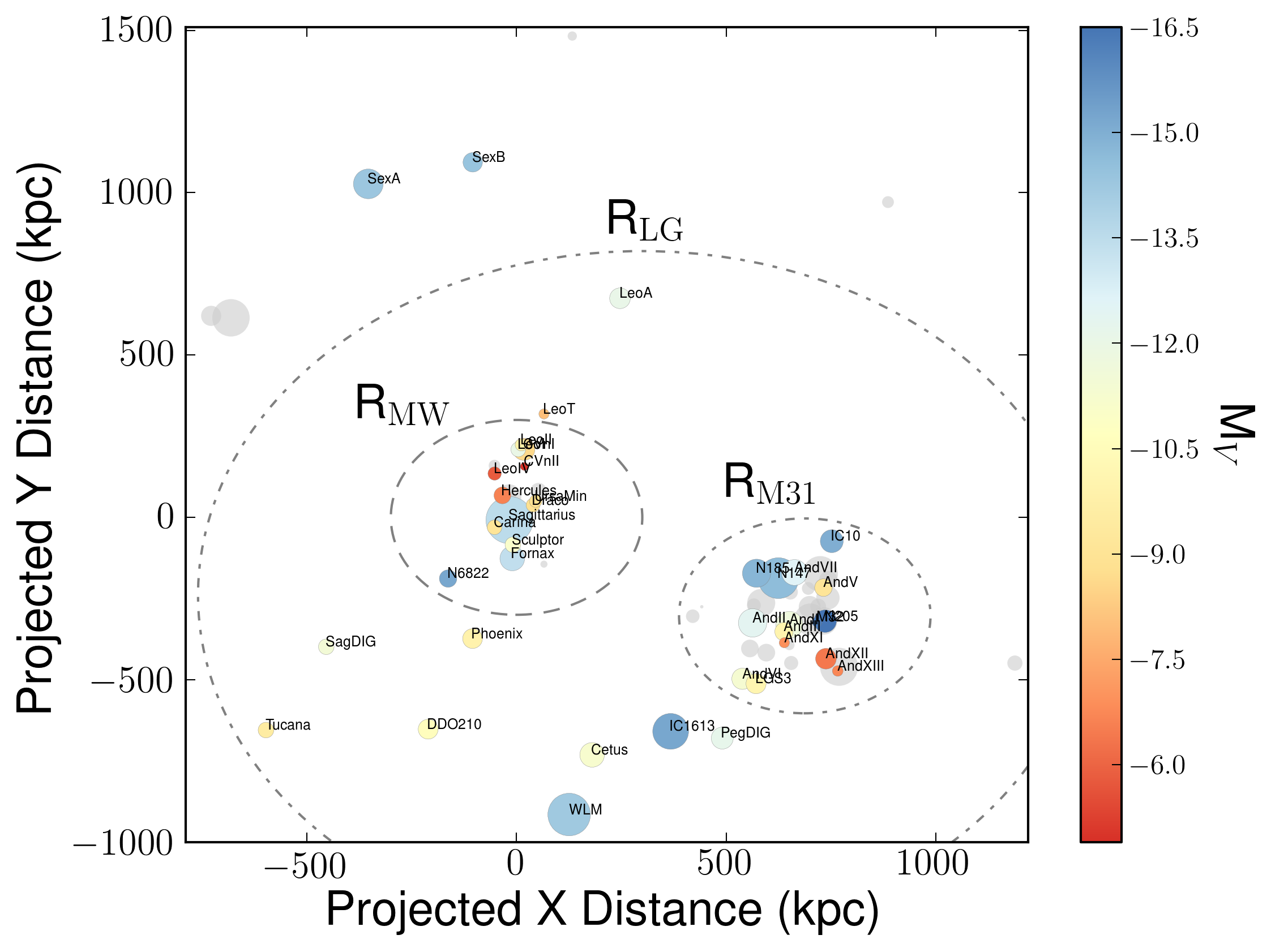}
\caption{The spatial distribution of LG dwarf galaxies projected into the supergalactic $X$-$Y$ plane.  Galaxies analyzed in this paper are color-coded by their absolute V-band magnitude, while other galaxies are left as grey points.  Outside the LG, the grey points are (from left to right): Antlia, NGC~3109, Leo~P, and UGC~4879. The size of each point is proportional to the galaxy's half-light radius.  Following \citet{mcconnachie2012} we adopt R$_{\mathrm{virial}} =$ 300 kpc for both the MW and M31, and a zero-velocity radius of 1060 kpc for the LG.}
\label{fig:lgdap_spatial_mv}
\end{center}
\end{figure*}

This paper is the first in a series on the SFHs of LG dwarf galaxies.  Here, we detail the process of translating CMDs into SFHs, including extensive exploration of the reliability and uncertainties in the measured SFHs, and explore a variety of science relating to the formation and evolution of the LG and low mass galaxies. We also tabulate all the data that will be used in future papers in the series.  The next paper in this series discusses the signatures of reionization in LG dwarf SFHs \citep{weisz2014b}, while future papers will explore a variety of SFH related science including quenching timescales and comparisons to state-of-the-art simulations of low mass systems.

This paper is organized as follows.  We summarize the sample selection and data reductions in \S \ref{sec:obs}.  We describe the process of measuring a SFH from a CMD and discuss a number of challenges related to CMD fitting and uncertainty analysis in \S \ref{sec:analysis}.  We proceed to empirically characterize the SFHs of the LG dwarf galaxies in \S \ref{sec:results}. In \S \ref{sec:sdss}, we compare our SFHs with spectroscopically based SFHs of star-forming SDSS galaxies and with SFH models based on abundance matching techniques. Finally, we provide a summary of our results in \S \ref{summary}.  Throughout this paper, the conversion between age and redshift assumes the Planck cosmology as detailed in \citet{planck2013}.

\section{The Data}
\label{sec:obs}

\subsection{The Sample}
\label{sec:sample}

For this paper, we have selected only dwarf galaxies that are located within the zero surface velocity of the LG \citep[$\sim$ 1 Mpc; e.g.,][]{vandenberg2000, mcconnachie2012}.  This definition excludes some dwarfs that have been historically associated with the LG, such as GR8 and IC~5152, but which are located well-beyond 1 Mpc. We have chosen to include two galaxies with WFPC2 imaging that are located on the periphery of the LG (Sex~A and Sex~B), because of their ambiguous association with the LG, the NGC~3109 sub-group, or perhaps neither \cite[although see][for discussion of the possible association of these systems]{bellazzini2013}.  Others in this sub-group, Antlia, UGC~4879, Leo~P, do not have WFPC2 imaging, and are therefore not included in this paper. Following previous LG galaxy atlases \citep[e.g.,][]{hodge1971, hodge1989, mateo1998, mcconnachie2012}, we have not included the SMC and LMC in this paper, as all the authors consider them to be too massive/luminous.  Instead, we are pursuing separate analysis of dozens of HST fields in these galaxies; preliminary results are presented in \citet{weisz2013b}.

Of the $\sim$ 60 known dwarfs located within $\sim$ 1 Mpc, we analyze those that have optical WFPC2 imaging in the HST archive.  Photometry for a majority of these galaxies was measured as part of the Local Group Stellar Photometry Archive\footnote{ \textcolor{blue}{\url{http://astronomy.nmsu.edu/logphot}}} \citep[LOGPHOT;][]{holtzman2006}, which provides a repository of uniformly reduced HST resolved star imaging of WFPC2 observations of LG targets taken during or prior to 2006.  After 2006, the majority of LG dwarf observations were taken with ACS and/or WFC3.  However, between the failure of ACS and completion of SM4, a handful of additional WFPC2 observations of LG dwarfs were executed.  From these observations, we have added 8 galaxies that were not reduced as part of LOGPHOT: Andromeda {\sc XI}, Andromeda {\sc XII}, Andromeda {\sc XIII}, Canes~Venatici {\sc I}, Canes~Venatici {\sc II}, Hercules, Leo {\sc IV}, Leo~T (HST-GO-11084; PI D.~Zucker); and one new central field for Sculptor (HST-GO-10888; PI A.~Cole).  To maintain uniformity throughout the sample, these datasets were reduced using the LOGPHOT photometric pipeline, and have been incorporated into the LOGPHOT online database.

\begin{deluxetable*}{cccccccccc}
\tabletypesize{\scriptsize}
\tablecolumns{10}
\tablewidth{0pt}
\tablecaption{Observational Properties of the WFPC2 Archival Data}
\tablehead{
\colhead{Galaxy} &
\colhead{Field} &
\colhead{Field RA} &
\colhead{Field DEC} &
\colhead{HST-PID} &
\colhead{Filters} &
\colhead{Exp. Times} &
\colhead{Comp. Limits} &
\colhead{Comp.} &
\colhead{No. Stars} \\
\colhead{Name} &
\colhead{ID} &
\colhead{(J2000)} &
\colhead{(J2000)} &
\colhead{} &
\colhead{} &
\colhead{(sec)} &
\colhead{(mag)} &
\colhead{Fraction} &
\colhead{in CMD} \\
\colhead{(1)} &
\colhead{(2)} &
\colhead{(3)} &
\colhead{(4)} &
\colhead{(5)} &
\colhead{(6)} &
\colhead{(7)} &
\colhead{(8)} &
\colhead{(9)} &
\colhead{(10)}
}
\startdata
Andromeda {\sc I} & u2e701 & 00 45 43 & +38 02 30 & 5325 & F450W,F555W &22020,12900 & 28.0,27.8 & 0.6,0.3 & 16580 \\
Andromeda {\sc II} & u41r01  & 01 16 22 & +33 25 45 & 6514 & F450W,F555W & 19500,8400 & 27.8,27.5 & 0.6,0.3 & 12801 \\
Andromeda {\sc III} & u56901  & 00 35 31 & +36 30 36 & 7500 & F450W,F555W & 20800,9600  & 27.5,27.3 & 0.6,0.3 & 6280 \\
Andromeda {\sc V} & u5c701  & 01 10 17 & +47 37 35 & 8272 & F450W,F555W & 11606,4820 & 27.3,27.1 & 0.6,0.3 & 3828 \\
Andromeda {\sc VI} & u5c703  & 23 51 46 & +24 34 56 & 8272 & F450W,F555W & 20800,8800  & 27.8,27.5 & 0.4,0.2 & 13684 \\
Andromeda {\sc VII} & u5c356  & 23 26 31 & +50 41 31 & 8192  & F606W,F814W & 600,600 & 25.6,24.5 & 0.4,0.7 & 7433 \\
Andromeda {\sc XI} & u9x701  & 00 46 20 & +33 47 29 & 11084  & F606W,F814W & 19200,26400 & 28.1,27.2 & 0.4,0.4 & 1174 \\
Andromeda {\sc XII} & u9x706  & 00 47 26 & +34 21 43 & 11084  & F606W,F814W & 19200,28800 & 28.1,27.2 & 0.3,0.4 & 534 \\
Andromeda {\sc XIII} & u9x712  & 00 51 50 & +33 00 18 & 11084  & F606W,F814W & 19200,31200 & 28.1,27.2 & 0.4,0.4 & 862 \\
Carina & u2lb01  & 06 41 48 & -50 58 10 & 5637 & F555W,F814W & 2400,2400 & 26.0,25.3 & 0.9,0.7 & 3277 \\
Cetus & u67201  & 00 26 14 & -11 02 26 & 8601 & F555W,F814W & 600,600 & 25.6,24.4 & 0.4,0.7 & 2498 \\
Canes~Venatici {\sc I} & u9x724 & 13 28 14 & +33 34 07 & 11084 & F606W,F814W & 4800,7200  & 27.0,26.1 & 0.6,0.6 & 1334 \\
Canes~Venatici {\sc II}& u9x722 & 12 57 09  & +34 19 09 & 11084 & F606W,F814W  & 3600,6000 & 27.0,26.1 & 0.5,0.6 & 900 \\
DDO~210 & u67203 & 20 46 51 & -12 51 29 & 8601 & F606W,F814W & 600,600 & 25.5,24.4 & 0.5,0.7 & 3684 \\
Draco & u2oc01 & 17 20 10 & +57 54 27 & 6234 & F606W,F814W & 2200,2600  & 26.2,25.4 & 0.8,0.8 & 4000 \\
Fornax & u2lb02 & 02 40 07 & -34 32 19 & 5637 & F555W,F814W & 2400,2400 & 26.6,25.6 & 0.4,0.7 & 18620 \\
Hercules & u9x723 & 16 31 01 & +12 47 59 & 11084 & F606W,F814W  & 3600,6000 & 27.0,26.1 & 0.6,0.6 & 802 \\
IC~10 & u40a01 & 00 20 25 & +59 17 00 & 6406 & F555W,F814W & 14000,14000 & 26.7,25.3 & 0.3,0.2 & 29298 \\ 
IC~1613 & u40401 & 01 04 48 & +02 07 06 & 6865 & F555W,F814W & 10700,10700 & 27.4,26.5 & 0.3,0.5 & 42496 \\
IC~1613 & u56750 & 01 04 26 & +02 03 16 & 7496 & F555W,F814W & 19200,38400 & 27.6,27.0 & 0.5,0.6 & 16816 \\
Leo~A & u2x501 & 09 59 26 & +30 44 21 & 5915 & F555W,F814W & 1800,1800  & 26.2,25.4 & 0.4,0.6 & 7655 \\
Leo~A & u64701 & 09 59 33 & +30 43 39 & 8575 & F555W,F814W & 8300,8300  & 27.2,26.2 & 0.2,0.4 & 4029 \\
Leo~{\sc I} & u27k01 & 10 08 26 & +12 18 33 & 5350 & F555W,F814W & 6050,5100  & 26.6,25.6 & 0.4,0.8 & 38358 \\
Leo~{\sc II} & u28v01 & 11 13 32 & +22 09 30 & 5386 & F555W,F814W & 4960,4960  & 26.9,25.9 & 0.5,0.7 & 14406 \\
Leo~{\sc IV} & u9x721 & 11 32 56 & -00 33 10 & 11084 & F606W,F814W & 3600,6000 & 26.8,25.9 & 0.6,0.6 & 638 \\
Leo~T & u9x717 & 09 34 52 & +17 03 09 & 11084 & F606W,F814W & 19200,32100 & 27.5,26.9 & 0.6,0.6 & 3847 \\
LGS~3 & u58901 & 01 03 52 & +21 53 06 & 6695 & F555W,F814W & 9200,9200 & 27.2,26.2 & 0.5,0.7 & 4014 \\
M32 & u28808 & 00 43 04 & +40 54 33 & 5233 & F555W,F814W & 2000,2000  & 26.3,25.3 & 0.5,0.5 & 48284 \\
NGC~147 & u2ob01 & 00 33 11 & +48 30 32 & 6233 & F555W,F814W & 13200,7800  & 27.6,26.2 & 0.3,0.5 & 37354 \\
NGC~185 & u3kl02 & 00 38 48 & +48 18 47 & 6699 & F555W,F814W & 2600,8400 & 26.6,26.2 & 0.4,0.2 & 38871 \\
NGC~185 & u3kl03 & 00 39 06 & +48 20 00 & 6699 & F555W,F814W & 2600,8400  & 26.2,25.8 & 0.5,0.2 & 41484 \\
NGC~185 & u3kl04 & 00 39 16 & +48 22 54 & 6699 & F555W,F814W & 2600,8400  & 26.7,26.3 & 0.5,0.2 & 24536 \\
NGC~205 & u3kl06 & 00 40 33 & +41 39 40 & 6699 & F555W,F814W & 2600,8100  & 26.2,25.8 & 0.4,0.2 & 44518 \\
NGC~205 & u3kl09 & 00 40 25 & +41 42 25 & 6699 & F555W,F814W & 2600,8100  & 26.2,25.8 & 0.2,0.1 & 40135 \\
NGC~205 & u3kl10 & 00 39 53 & +41 47 19 & 6699 & F555W,F814W & 2600,8100  & 26.4,26.0 & 0.6,0.3 & 26747 \\
NGC~6822 & u37h02 & 19 44 50 & -14 51 34 & 6813 & F555W,F814W & 3900,3900 & 26.7,25.5 & 0.3,0.4 & 34544 \\
NGC~6822 & u37h04 & 19 44 57 & -14 48 49 & 6813 & F555W,F814W & 3900,3900 & 26.1,25.4 & 0.5,0.4 & 50491 \\
NGC~6822 & u37h05 & 19 45 00 & -14 43 07 & 6813 & F555W,F814W & 3900,3900 & 26.7,25.5 & 0.3,0.6 & 31174 \\
Pegasus & u2x504 & 23 28 33 & +14 44 18 & 5915 & F555W,F814W & 1800,1800 & 26.1,25.1 & 0.6,0.7 & 18545 \\
Phoenix & u64j01 & 01 51 07 & -44 26 40 & 8706 & F555W,F814W & 7200,9600 & 27.1,26.2 & 0.3,0.6 & 22464 \\
Phoenix & u64j03 & 01 51 08 & -44 24 04 & 8706 & F555W,F814W & 11900,14300 & 27.2,26.4 & 0.6,0.7 & 9093 \\
Sag~DIG & u67202 & 19 29 56 & -17 41 07 & 8601 & F606W,F814W & 600,600  & 25.5,24.4 & 0.4,0.7 & 4078 \\
Sagittarius & u37501 & 18 55 05 & -30 16 39 & 6614 & F555W,F814W & 2160,1800  & 26.1,24.9 & 0.5,0.4 & 5741 \\
Sagittarius & u37503 & 19 06 12 & -30 25 46 & 6614 & F555W,F814W & 2160,1800  & 26.1,24.9 & 0.4,0.4 & 3383 \\
Sculptor & u43502 & 01 00 07 & -33 27 34 & 6866 & F555W,F814W  & 5000,5200 & 27.0,26.0 & 0.7,0.6 & 1016 \\
Sculptor & u9xm02 & 01 00 08 & -33 42 37 & 10888 & F450W,F814W  & 4800,7200 & 26.0,25.5 & 0.8,0.3 & 7527 \\
Sex~A & u2x502 & 10 11 00 & -04 42 14 & 5915 & F555W,F814W & 1800,1800 & 26.4,25.3 & 0.2,0.5 & 17682 \\
Sex~A & u56710 & 10 10 56 & -04 41 30 & 7496 & F555W,F814W & 19200,38400 & 27.6,27.0 & 0.4,0.5 & 38578 \\
Sex~B & u67204 & 10 00 02 & +05 19 31 & 8601 & F606W,F814W & 600,600 & 25.5,24.4 & 0.2,0.5 & 13379 \\
Tucana & u2cw02 & 22 41 44 & -64 25 36 & 5423 & F555W,F814W  & 7200,14400 & 27.5,26.8 & 0.6,0.5 & 11355 \\
Ursa~Minor & u2pb01 & 15 08 33 & +67 12 15 & 6282 & F555W,F814W & 2300,2500 & 26.2,25.4 & 0.8,0.6 &  2194 \\
WLM & u37h01 & 00 01 49 & -15 28 01 & 6813   & F555W,F814W   & 5300,5300  & 27.3,26.4 & 0.3,0.4 & 12191 \\
WLM & u48i01 & 00 01 58 & -15 26 02 & 6798 & F555W,F814W   & 4900,4900 & 26.6,25.7 & 0.3,0.5 & 28356
\enddata
\tablecomments{The observational properties of each field analyzed in this paper.  The corresponding CMDs for each field are shown in Figure \ref{fig:cmdgrid}. Column (2) is the unique HST field identifier.  Column (9) indicates the completeness fraction at the adopted faint photometric limit of our CMDs as listed in in Column (8).}

\label{tab1}
\end{deluxetable*}

In total, our sample consists of 53 fields located in 40 LG dwarf galaxies, or roughly 2/3 of the known LG dwarf population.  Their spatial distribution is shown in Figure \ref{fig:lgdap_spatial_mv}.  Galaxies in our sample span a wide range in luminosity, mean metallicity, and morphological type ranging from extremely low luminosity metal- and gas-poor systems such as Canes~Venatici {\sc II} (M$_{V} = -$4.9) to more luminous, higher metallicity, and gas-rich galaxies such as NGC~6822 (M$_{V} = -$15.2).  The tidally disrupted galaxy Sagittarius dwarf is also included.   Following conventional morphological designations, our sample consists of 23 dwarf spheroidals (dSphs), 4 dwarf ellipticals (dEs\footnote{We include M32 in this category for convenience, even though it is a compact elliptical \citep[e.g.,][]{kormendy1985, kormendy2012}}), 8 dwarf irregulars (dIs), and 5 transition dwarfs (dTrans). The latter class are galaxies that have observed \hi\ but have no detectable \halpha\ emission \citep[e.g.,][]{mateo1998}, and have historically been associated with the transition from gas-rich to gas-poor \citep[but see][for discussion of the various processes that can produce dTrans galaxies]{weisz2011a}.

The majority of our sample consists of single WFPC2 pointings targeting the central, highest surface brightness region of each galaxy.  In some cases, there are multiple pointings available per galaxy.  We evaluated the quality of CMDs of the additional fields (e.g., filter combinations, photometric depth), and added fields that were likely to help construct a more globally representative SFH of that system.  The final selection of galaxies and fields, along with their observational characteristics, are listed in Table \ref{tab1}. 

Because our sample was selected based on availability of WFPC2 observations, there are at least three important selection effects we must consider.  First, our sample is neither complete nor unbiased (in any sense) in its construction (e.g., more luminous dwarfs tend to have been discovered and observed with HST first, whereas a number of LG dwarfs have no WFPC2 imaging).  Second, the fields typically cover variable fractional areas of of each galaxy.  For the closest MW companions, WFPC2 fields typically subtend a small fraction ($<$ 10\%) of the galaxy.  In contrast, for galaxies on the edge of the LG, the WFPC2 footprints cover a larger physical area of the galaxy ($>$ 30\%); the majority of galaxies fall in between these two extremes.  Third, field placement varies from galaxy to galaxy.  In some cases, the fields were centered on a galaxy, in others, off-center locations were chosen to minimize crowding effects.  

The net result is that the selection function of our sample is quite complicated.  While we do not attempt to quantitatively model the potential selection biases, we do qualitatively discuss them in cases where the heterogeneity of the data may compromise any physical interpretation of our results.  We have summarized the spatial locations and areal coverage of each field in Table \ref{tab2}, and list other observable properties in Table \ref{tab:alan_data}.

\begin{figure*}
\begin{center}
\plotone{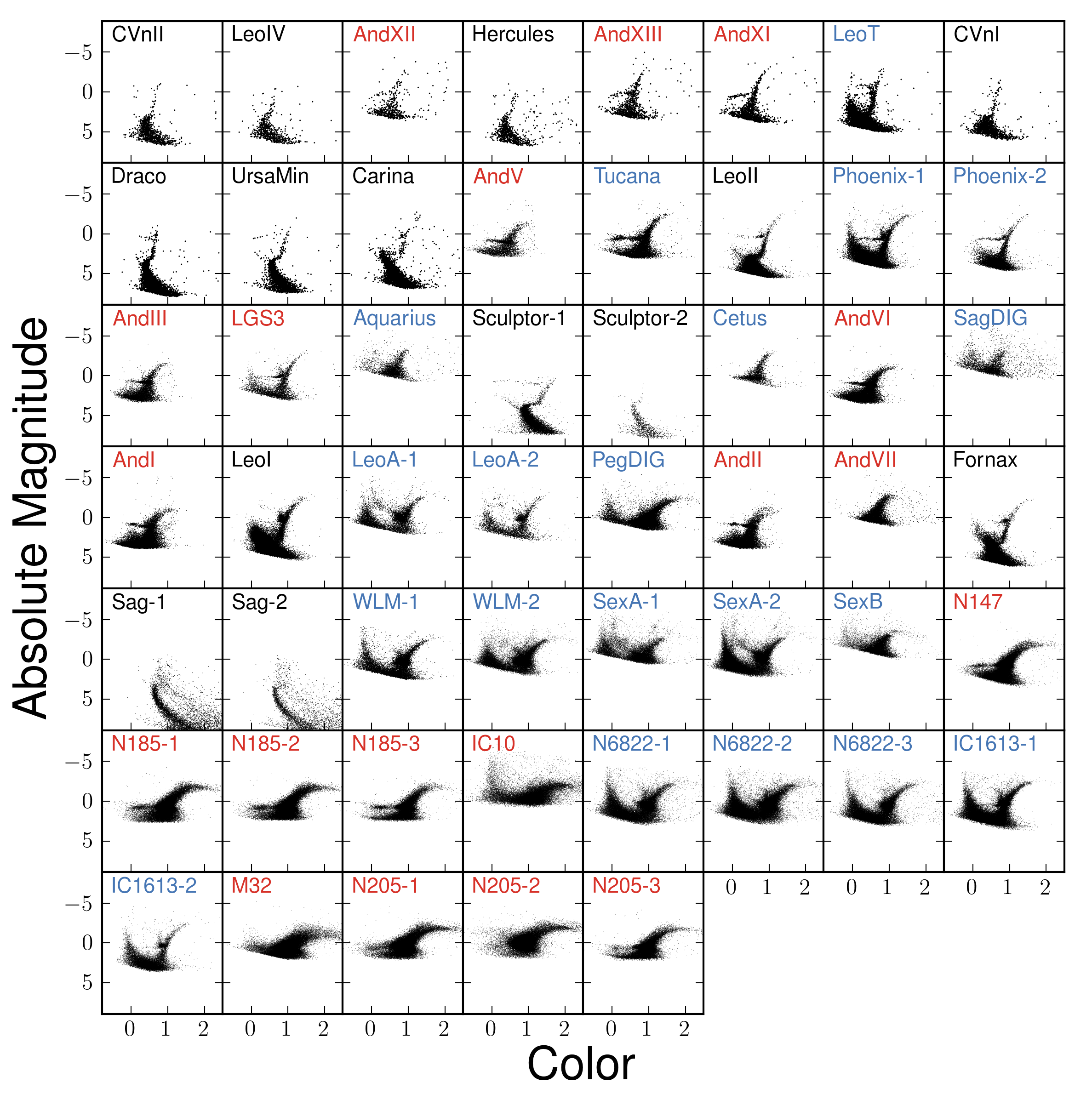}
\caption{CMDs of all fields analyzed in this paper, sorted by the increasing absolute V-band magnitude. The field names are color-coded by a galaxy's association with the MW (black), M31 (red), or the field (blue). Given the heterogeneity in filter sets, we have indicated only a canonical color and reddest available magnitude on the $x$ and $y$ axes, respectively.  The specific filter combinations are listed in Table \ref{tab1}. All galaxies associated with the MW have CMDs that extend below the oldest MSTO.  However, only two field galaxies (Phoenix and Leo~T) and no galaxies associated with M31 have comparably deep CMDs. }
\label{fig:cmdgrid}
\end{center}
\end{figure*}

\subsection{Photometry}
\label{sec:photometry}

The flux of any stellar-like objects in the images were measured using point spread function (PSF) photometry with the \texttt{HSTPHOT} software package, which has been specifically designed to handle the under-sampled PSF of WFPC2 images \citep{dolphin2000b}.  The majority of the photometry was derived as part of LOGPHOT \citep{holtzman2006}, and the resulting datasets were downloaded from the LOGPHOT online database.  For the 8 additional galaxies, we ran the LOGPHOT pipeline on the \texttt{c0f} and \texttt{c1f} WFPC2 images that were downloaded from the HST archive, ensuring homogeneity in data reduction and calibration.  Following the data quality recommendations in \citet{holtzman2006}, we designated good stars to be sources with data flag values of 0 or 1. For each pointing we also ran or acquired from the LOGPHOT  at least $\sim$ 1.2$\times$10$^{5}$ artificial star tests, which provide an accurate means of modeling the noise and completeness of the observed CMDs.  More details of the entire WFPC2 photometric reduction process can be found in \citet{dolphin2000b} and \citet{holtzman2006}.  

For a minority of the galaxies, we made small modifications to the photometric catalogs.  In the cases of Fornax u2lb02 and WLM u37h01, we excluded WFPC2 chips that were centered on globular clusters, to reduce potential age biases in the SFHs.  In the case of Tucana, we combined two adjacent WFPC2 fields (u2cw01 and u2cw02) into single photometry and false star lists, after verifying that their completeness functions were statistically identical.  Finally, in the case of Sagittarius, we combined a handful of overlapping fields, after removing duplicate stars, and utilized off galaxy pointings as the background field.  Specifically, fields u37502 and u37504 were concatenated to fields u37501 and u37503, respectively.  Fields u37505 and u37506 were used as empirical proxies for modeling contamination due to intervening Milky Way populations.  The data for all fields discussed here are available via the LOGPHOT online repository.

\subsection{Color-Magnitude Diagrams}
\label{sec:CMDs}

In Figure \ref{fig:cmdgrid}, we plot the CMDs for all fields in our sample in order of increasing V-band luminosity.  Because of the variety of filter combinations, the $x$- and $y$-axes simply indicate a canonical absolute magnitude (based on distance and extinction determinations described in \S \ref{sec:distances}) and color. The specific filter combinations for any particular field are listed in Table \ref{tab1}.  The magnitude plotted on the $y$-axis is the reddest filter available for a given field.

The archival WFPC2 data generally yield CMDs that reach below the oldest MSTO out to a distance of $\sim$ 450 kpc, i.e., 1.5R$_{\mathrm{virial}}$ \citep[R$_{\mathrm{virial}}$ $\sim$ 300 kpc;][]{mcconnachie2012}.  As shown in Figure \ref{fig:lgdap_spatial_mv}, this includes all MW satellites and two galaxies, Phoenix and Leo~T, that are not considered satellites. 

The CMDs shown in Figure \ref{fig:cmdgrid} are diverse.  The faintest galaxies typically have sparsely populated and predominantly older sequences, e.g., the red giant branch (RGB), horizontal branch (HB), and sub-giant branch (SGB) --- with the notable exception of Leo~T, which has a clear but poorly populated young blue plume, despite being among the faintest systems.  However, among the faintest systems, the CMD morphologies vary considerably.  First, the width of the RGB and/or SGB of the faintest systems can sometimes be larger than the photometric errors (typically $< $0.05-0.1 mag in color), suggesting that not all faint galaxies are mono-metallic and/or single age populations.  Second, the faintest Andromeda satellites in our sample (And {\sc XI, XII, XIII}) all show well-populated, and red HB populations compared to MW satellites of comparable luminosities.  This finding is similar to previous comparisons of HB morphologies between M31 and MW dwarfs \citep[e.g.,][who suggested the possibility of systematic age differences between satellites of the two systems]{dacosta2002}.  We discuss potential differences in the SFHs of MW and M31 satellites in \S \ref{sec:masstrends}.

Some of the slightly brighter, intermediate luminosity dwarfs (i.e., Draco to Fornax in Figure \ref{fig:cmdgrid}) share similar characteristics with the faintest galaxies, e.g., predominantly old populations and sparsely populated blue plumes.  However, with increasing luminosity, the CMD morphologies change to have broader RGBs and/or SGBs, clear populations of asymptotic giant branch (AGB) stars, and clear HB populations that exhibit varying types of morphologies and densities.  Some of the intermediate luminosity dwarfs also have clear blue plume populations.   In some cases, the blue plume populations are believed to be intermediate to young aged MS stars (e.g., Leo {\sc II}, Phoenix, Carina, LGS~3), whereas in others (e.g., Draco, Sculptor) these populations are thought to be older blue stragglers \citep[e.g.,][]{mapelli2007, momany2007, mapelli2009, monelli2012, sand2010, santana2013}. The dIrrs in this luminosity range (e.g., Leo~A, Sag~DIG) show clear young blue plume populations. 

Finally, the most luminous systems in our sample (i.e., brighter than WLM) are either dEs associated with the M31 subgroup or dIrrs.  The CMD morphologies of the most luminous dIrrs are similar to those in the intermediate luminosity range, only with more strongly populated sequences.  However, the dEs have no clear analogs among lower luminosity systems.  While they appear to be predominantly old like dSphs, the dEs have very broad and red RGBs, indicative of high metallicites and/or multiple generations of star formation over their lifetimes.  The luminous dIrrs and dEs have CMDs that typically do not extend more than a magnitude below the HB, chiefly because of their large distances and high surface brightnesses, both of which increase crowding effects.

\section{Analysis}
\label{sec:analysis}

\subsection{Measuring the Star Formation Histories}
\label{sec:measuresfhs}

We measured the SFH of each field using the maximum likelihood CMD fitting routine, \texttt{MATCH} \citep{dolphin2002}.  Briefly, \texttt{MATCH} works as follows:  it accepts a range of input parameters (e.g., IMF slope, binary fraction, age and metallicity bin widths), uses these parameters to construct synthetic CMDs of simple stellar populations (SSPs), and then linearly combines them with a model foreground CMD to form a composite model CMD with a complex SFH.    The composite model CMD is then convolved with the noise model from the artificial star tests (i.e., completeness, photometric uncertainties, and color/magnitude biases).  The resulting model CMD is then compared to the observed CMD using a Poisson likelihood statistic.  This process is repeated for various weights on the SSPs, until the overall likelihood function is maximized.  The synthetic CMD that best matches the observed CMD is deemed to have the most likely SFH that produced the data.  More technical details of \texttt{MATCH} are presented in \citet{dolphin2002}. 

In this paper, we have adopted the following parameters for the SFH solutions:  a Kroupa IMF \citep{kroupa2001}, a binary star fraction of 0.35 with the mass of the secondary drawn from a uniform distribution of mass ratios, and the Padova stellar evolution models of \citep{girardi2010}, with mass limits ranging from 0.15 to 120 \msun.  We have selected the Padova models for our analysis as they are currently the only stellar models that span the full range in mass, metallicity, and evolutionary phase needed to analyze this diverse collection of galaxies.  As described in \S \ref{sec:sfherrors}, we also characterize uncertainties in the SFHs due to our choice in stellar models, which helps to mitigate known shortcomings present in many stellar evolution libraries \citep[c.f.,][]{gallart2005}.  

To balance computational speed with reasonable time resolution, we have selected 50 time bins that are logarithmically spaced by 0.1 dex from $\log(t)$ $=$ 6.6 to 8.7, and by 0.05 dex from $\log(t)$ $=$ 8.7 to 10.15.  We allowed the program to search for metallicities between  $[M/H]$ $=$ $-$2.3 and 0.1 with a resolution of 0.1 dex.  However, due to the variable depth of the data across the sample and difficulty in constraining chemical evolution models solely from optical broadband CMDs, we required the mean metallicity to  increase monotonically with time.  Within each time bin, however, we permitted a 1-$\sigma$ metallicity dispersion of 0.15 dex, which provides the flexibility to capture potential metallicity spreads at each age.  For consistency across the sample, we imposed this metallicity restriction on all fields, even those with CMDs that extend below the oldest MSTO.  Through extensive testing, we found that for such deep CMDs the resulting SFHs and chemical evolution models were consistent whether the condition of monotonic chemical evolution was required or not.

\begin{deluxetable*}{ccccccccc}
\tablecaption{Observational Properties of each WFPC2 Field}
\tablehead{
\colhead{Galaxy} &
\colhead{Field} &
\colhead{(M-m)$_0$} &
\colhead{A$_{\mathrm{V}}$}&
\colhead{R$_{\mathrm{GC}}$}&
\colhead{$\frac{\mathrm{A}_{\mathrm{WFPC2}}}{\pi\mathrm{r}^2_{\mathrm{h}}}$} &
\colhead{Completeness Limits} &
\colhead{ $\sigma(\log(\mathrm{T_{eff}}))$} &
\colhead{$\sigma(\log(\mathrm{M_{bol}}))$} \\
\colhead{} &
\colhead{} &
\colhead{} &
\colhead{} &
\colhead{(pc)} &
\colhead{} &
\colhead{(absolute mag)} &
\colhead{} &
\colhead{} \\
\colhead{(1)} &
\colhead{(2)} &
\colhead{(3)} &
\colhead{(4)} &
\colhead{(5)} &
\colhead{(6)} &
\colhead{(7)} &
\colhead{(8)} &
\colhead{(9)} 
}
\startdata
Andromeda {\sc I} & u2e701 & 24.49 & 0.15 & 140 & 0.24 & 3.32, 3.16 & 0.013 & 0.27 \\
Andromeda {\sc II} & u41r01  & 24.09 & 0.17 & 330 & 0.06 & 3.50, 3.24 & 0.013 & 0.27 \\
Andromeda {\sc III} & u56901  & 24.30 & 0.15 & 200 & 0.48 & 3.01, 2.85 & 0.015 & 0.34 \\
Andromeda {\sc V} & u5c701  & 24.49 & 0.34 & 23 & 1.2 & 2.38, 2.26 & 0.015 & 0.27 \\
Andromeda {\sc VI} & u5c703  & 24.60 & 0.17 & 16 & 0.44 & 2.99, 2.73 & 0.015 & 0.34 \\
Andromeda {\sc VII} & u5c356  & 24.58 & 0.46 & 220 & 0.19 &  0.60, $-$0.36 & 0.017 & 0.19 \\
Andromeda {\sc XI} & u9x701  & 24.35 & 0.36 & 220 & 4.6 & 3.42, 2.63 & 0.015 & 0.33 \\
Andromeda {\sc XII} & u9x706  & 24.44 & 0.31 & 201 & 1.6 & 3.37, 2.57 & 0.020 & 0.33 \\
Andromeda {\sc XIII} & u9x712  & 24.57 & 0.36 & 56 & 1.1 & 3.20, 2.41 & 0.020 & 0.33 \\
Carina & u2lb01  & 20.06 & 0.17 & 57 & 0.03 & 5.76, 5.14 & 0.013 & 0.19 \\
Cetus & u67201  & 24.39 & 0.08 & 170 & 0.23 & 1.14, $-$0.03  & 0.019 & 0.18 \\
Canes~Venatici {\sc I} & u9x724 & 21.75 & 0.04 & 150 & 0.03 & 5.21, 4.32 & 0.013 & 0.19 \\
Canes~Venatici {\sc II} & u9x722 & 20.95 & 0.03 & 14 & 0.92 & 5.02, 5.13 & 0.013 & 0.19 \\
DDO~210 & u67203 &24.85 & 0.14 & 200 & 1.08 & 0.52, $-$0.54 & 0.026 & 0.35 \\
Draco & u2oc01 & 19.54 & 0.07 & 12 & 0.02 & 6.60, 5.82 & 0.013 & 0.19 \\
Fornax & u2lb02 & 20.79 & 0.06 & 240 & 0.01 & 5.75, 4.77 & 0.013 & 0.19 \\
Hercules & u9x723 & 20.70 & 0.17 & 21 & 0.03 & 6.14, 5.29 & 0.013 & 0.19 \\
IC~10 & u40a01 & 24.44 & 2.33 & 360 & 0.33 & $-$0.13, $-$0.57 & 0.026 & 0.35 \\
IC~1613 & u40401 & 24.28 & 0.07 & 13 & 0.05 & 3.04, 2.18 & 0.020 & 0.31 \\
IC~1613 & u56750 & 24.48 & 0.07 & 1400 & 0.05 & 3.25, 2.68 & 0.020 & 0.31 \\
Leo~A & u2x501 & 24.52 & 0.06 & 150 & 0.5 & 1.62, 0.84 & 0.021 & 0.16 \\
Leo~A & u64701 & 24.52 & 0.06 & 410 & 0.5 & 2.62, 1.64 & 0.020 & 0.27 \\
Leo~{\sc I} & u27k01 & 21.95 & 0.10 & 40 & 0.2 & 4.55, 3.59 & 0.013 & 0.19 \\
Leo~{\sc II} & u28v01 & 21.56 & 0.05 & 57 & 0.34 & 5.29, 4.31 & 0.013 & 0.19 \\
Leo~{\sc IV} & u9x721 & 20.85 & 0.07 & 53 & 0.11 & 5.89, 5.01 & 0.013 & 0.19 \\
Leo~T & u9x717 & 23.15 & 0.09 & 46 & 2.37 & 4.26, 3.69 & 0.013 & 0.19 \\
LGS~3 & u58901 & 23.96 & 0.11 & 160 & 0.53 & 3.13, 2.18 & 0.020 & 0.31 \\
M32 & u28808 & 24.59 & 0.17 & 310 & 11. & 1.54, 0.61 & 0.021 & 0.16 \\
NGC~147 & u2ob01 & 24.41 & 0.37 & 36 & 0.23 & 2.81, 1.56 & 0.020 & 0.27 \\
NGC~185 & u3kl02 & 24.09 & 0.41 & 400 & 0.41 & 2.08, 1.86 & 0.018 & 0.18 \\
NGC~185 & u3kl03 & 24.09 & 0.46 & 240 &0.41 & 1.64, 1.43 & 0.018 & 0.18 \\
NGC~185 & u3kl04 & 24.09 & 0.55 & 720 & 0.41 & 2.05, 1.87 & 0.018 & 0.18 \\
NGC~205 & u3kl06 & 24.54 & 0.17 & 600 & 0.38 & 1.49, 1.16 & 0.021 & 0.16 \\
NGC~205 & u3kl09 & 24.54 & 0.17 & 340 & 0.38 & 1.49, 1.16 & 0.021 & 0.16 \\
NGC~205 & u3kl10 & 24.54 & 0.17 & 2000 & 0.38 & 1.69, 1.36 & 0.018 & 0.18 \\
NGC~6822 & u37h02 & 23.35 & 0.85 & 600 & 0.33 &2.48, 1.62 & 0.020 & 0.27 \\
NGC~6822 & u37h04 & 23.35 & 1.01 & 200 & 0.33 &1.71, 1.43 & 0.018 & 0.18 \\
NGC~6822 & u37h05 & 23.35 & 0.73 & 580 & 0.33 & 2.60, 1.70 & 0.020 & 0.27 \\
Pegasus & u2x504 & 24.94 & 0.19 & 230 & 0.53 & 0.97, 0.05 & 0.019 & 0.18 \\
Phoenix & u64j01 & 23.10 & 0.04 & 80 & 0.16 & 3.97, 3.09 & 0.013 & 0.19 \\
Phoenix & u64j03 & 23.10 & 0.04 & 320 & 0.16 & 4.07, 3.29 & 0.013 & 0.19 \\
Sag~DIG & u67202 & 25.11 & 0.34 & 300 & 2.81 & 0.08, $-$0.92 & 0.026 & 0.35 \\
Sagittarius & u37501 & 17.32 & 0.46 & 120 & $<$ 0.01 & 8.31, 7.30 & 0.013 & 0.19 \\
Sagittarius & u37503 & 17.32 & 0.35 & 1080 & $<$ 0.01 & 8.42, 7.37 & 0.013 & 0.19 \\
Sculptor & u43502 & 19.56 & 0.05 & 380 & 0.02 & 7.39, 6.41 & 0.013 & 0.19 \\
Sculptor & u9xm02 & 19.56 & 0.05 & 8 &  0.02 & 6.38, 5.91 & 0.013 & 0.19 \\
Sex~A & u2x502 & 25.72 & 0.12 & 290 & 0.38 & 0.56, $-$0.49 & 0.017 & 0.19 \\
Sex~A & u56710 & 25.72 & 0.12 & 500 & 0.38 & 1.76, 1.21 & 0.021 & 0.16 \\
Sex~B & u67204 & 25.70 & 0.09 & 260 & 2.06 & $-$0.29, $-$1.36 & 0.024 & 0.41 \\
Tucana & u2cw02 & 24.73 & 0.09 & 200 & 1.92 & 2.67, 2.01 & 0.020 & 0.27 \\
Ursa~Minor & u2pb01 & 19.31 & 0.09 & 80 & 0.03 & 6.80, 6.03 & 0.013 & 0.19 \\
WLM & u37h01 & 24.89 & 0.10 & 600 & 0.16 & 2.31, 1.45 & 0.018 & 0.18 \\
WLM & u48i01 & 24.89 & 0.10 & 440 & 0.16 & 1.61, 0.75 & 0.021 & 0.16 
\enddata
\tablecomments{The adopted values used for SFH analysis in each fields.  Measurement of the adopted distance and extinction values are discussed in \S \ref{sec:distances}.  We have not listed uncertainties in either distance or extinction because they contribute a negligible amount to the total SFH error budget, as demonstrated in \citet{dolphin2013}.  Column (5) lists the projected distance between the galaxy coordinates listed in \citet{mcconnachie2012} and the center of each WFPC2 field.  Column (6) is the ratio of the WFPC2 field area to a circular area defined by the galaxy's half-light radius.  Column (7) lists the absolute value of the completeness limits from Table \ref{tab1} and uses the distance and extinction values from Columns (2) and (3).  Columns (8) and (9) list the effective temperature and bolometric magnitude shifts applied to the best fit SFH model in order to mimic uncertainties in the stellar models used to derive the SFHs, i.e., the systematic uncertainties.  The details of systematic uncertainty analysis as described in \citet{dolphin2012}.}
\label{tab2}
\end{deluxetable*}

For each field, we adopted the photometric limits indicated in Table \ref{tab1}.  These limits generally fall within the 30-50\% completeness limits of the data.  In the case of more crowded fields, e.g., NGC~205, DDO~210, we adopted more liberal limits (i.e., 30-50\%), for maximum photometric depth, whereas for deeper, less crowded fields (e.g., Sculptor, Draco), we generally adopted limits closer to 60-70\% completeness, as these conservative limits are typically below the oldest MSTO those CMDs.  We examine the issue of CMD depth and SFH recovery further in \S \ref{sec:lcidcompare}.

We additionally modeled intervening Milky Way foreground stars by adding model foreground stars to the composite synthetic CMD.  Specifically, we generated synthetic CMDs using the Dartmouth stellar evolution libraries \citep{dotter2008} and the Milky Way structural model presented in \citet{dejong2010b}, and allowed \texttt{MATCH} to linearly scale the resulting CMD to model any putative foreground stars that may have contaminated each CMD.  The Dartmouth models extend to slightly lower stellar masses (0.1 \msun) than the Padova models (0.15 \msun), which allow for a more realistic foreground population model.  In general, we found the resulting scaling factors to be quite small, in agreement with the prior expectation that few foreground stars would fall into a typical HST field of view.  As previously discussed, in the case of Sagittarius dSph, we used off-galaxy HST/WFPC2 fields as the empirical foreground CMD in the modeling process.

We have not explicitly accounted for blue straggler stars when measuring the SFHs.  The optical colors and magnitudes of blue stragglers are the same as $\sim$ 1-3 Gyr main sequence stars.  However, because there are currently no generative models for blue stragglers, it is not possible to explicitly include them in the SFH measurement process.  It is possible to simply exclude putative blue stragglers from the CMD fitting process \textit{a priori}.  However, it is not always clear whether stars in the blue plume of early type galaxies are blue stragglers or main sequence stars \citep[e.g.,][]{martin2008, sand2010, okamoto2012}.  If they are main sequence stars, excluding them would remove critical information about interesting late time star formation in early type galaxies.  Thus, we adopt the approach of including the blue plume in every field, and use literature studies as guidance for interpreting the SFHs.  Several studies have empirically shown that blue plume stars in early type galaxies comprise 5-7\% of the system's total stellar mass \citep[e.g.,][]{mapelli2007, momany2007, mapelli2009, sand2009, sand2010, sand2012, monelli2012, santana2013}, making them minor contributors to the overall SFH.  We approach any science interpretations that involve the late time SFHs in early type galaxies with the appropriate caution.

Finally, for star-forming dwarfs we included an age-dependent differential extinction model for stars with ages of $<$ 100 Myr as specified in \citet{dolphin2003}.  In this model, stars with ages between 100 and 40 Myr have a linearly increasing maximum differential extinction from 0.0 to 0.5 mag in A$_{V}$ and stars $<$ 40 Myr in age all have a maximum extinction of 0.5 mag in A$_{V}$.  This model has been shown to provide optimal fits to the young stellar populations in a variety of nearby dwarf galaxies \citep[e.g.,][]{dolphin2003,skillman2003,weisz2011a}, and is necessary to accurately reproduce the observed integrated blue-optical and ultra-violet light of dwarf galaxies in the Local Volume \citep{johnson2013}.

\subsection{Adopted Distance and Extinction Values}
\label{sec:distances}

When computing the most likely SFHs, we adopted single values for distance and foreground Milky Way extinction.  Uncertainties in the SFHs due to precision in distance and extinction are a negligibly small part of the total SFH error budget; both random uncertainties and systematic effects due to the underlying stellar models can be over an order of magnitude larger \citep[e.g.,][]{dolphin2012, dolphin2013}.  We determined the single values for both distance and extinction through the following procedure:

\begin{enumerate}

\item Calibrate the distance indicators: For CMDs with a well-populated tip of the red giant branch (TRGB), we used maximum likelihood fitting technique from \citet{makarov2006} to measure the magnitude of the TRGB. We compared each distance with the apparent F814W distance modulus obtained by allowing MATCH to find the distance and extinction from the best fit CMD.  Comparing these two, we found the TRGB magnitude minus the apparent distance modulus of M$_{\mathrm{F814W}} = -$3.96.  We repeated the procedure for the HB and found the mean m$_{\mathrm{F555W}}$ HB magnitude minus the apparent distance modulus of M$_{\mathrm{F555W}} = +$0.66.

\item Measure apparent distances:  For all fields with a TRGB and/or HB measurement, we computed the apparent distance moduli using the calibration in the previous step.

\item Measure or adopt extinction:  For most fields, we adopted the \citet{schlafly2011} extinction values from NED at the location of the pointing.  For fields with A$_{\mathrm{V}} >$ 0.2, we used MATCH to determine the value of A$_{\mathrm{V}}$ that provided the best fit, given the apparent distance modulus.  

\item Adopt a distance:  For fields with both a TRGB and HB apparent distance measurement, we adopted the value that provided the best fit CMD.

\item For galaxies with multiple fields, we computed the absolute distance modulus per field, and set the final distance and extinction values to the average.
\item For galaxies with neither a strong HB nor a well-populated TRGB, we allowed \texttt{MATCH} to find the most likely distance after adopting the extinction value from \citet{schlafly2011}.  
\item For the Sagittarius dSph, we adopted the most likely \texttt{MATCH} distance and extinction combination, but required both fields to have the same distance.
\end{enumerate}

The adopted distance and extinction values for each field are listed in Table \ref{tab2}. We emphasize that our distance determinations are relative to the Padova stellar models, which we use both for CMD fitting and as a reference for HB and TRGB luminosities.

For comparison, we also have tabulated the distance moduli reported in \citet{mcconnachie2012} and the foreground extinction values from \citet{schlafly2011} in Table \ref{tab:alan_data}.  A cursory comparison between our distance and extinction values and those from the literature show good overall agreement.  However, there are a few examples of poor agreement.  The most extreme is for IC~10. Our derived extinction value for IC~10 is A$_{\mathrm{V}}=$2.3, while \citet{schlafly2011} find a value of A$_{\mathrm{V}}=$4.3.  However, IC~10 is located at low Galactic latitude and suffers from a high degree of differential extinction.  Therefore while the \citet{schlafly2011} value is an average in that direction of the sky, the HST observations targeted a low extinction window toward IC~10, which provides a qualitative reason for the difference.  Further, despite a 2 magnitude difference in A$_{\mathrm{V}}$, our derived distance modulus of 24.44 is in good agreement with the value of 24.50$\pm$0.12 reported by \citet{mcconnachie2012}.  There are also handful smaller disagreements ($\sim$ 0.1 dex) between our distance determinations and those in the literature. However, \citet{mcconnachie2012} emphasize the heterogenous nature of their listed distance measurements (i.e., they range from TRGB measurements to MS fitting) and caution that the reported uncertainties are only statistical, and therefore should be considered with lower limits.  Given these caveats, we find that our distance and extinction measurements are in good qualitative agreement with those commonly adopted in the literature.

\subsection{Characterizing Uncertainties in the Star Formation Histories}
\label{sec:sfherrors}

Uncertainties in SFH measurements can be broken into two sources: random and systematic.   Random uncertainties are due to the finite number of stars, and therefore generally scale in amplitude with the sparsity of the observed CMD.  In comparison, systematic uncertainties are due to inherent shortcomings with the stellar models (e.g., uncertainties in convective overshooting, mass loss) used to measure the SFHs, and generally scale as a function of photometric depth, i.e., deeper CMDs provide more secure leverage on the lifetime SFH of a galaxy relative to a shallower CMD because they contain more age sensitive features and older MSTOs. However, other factors such as the intrinsic SFH and morphologies of key CMD features (e.g., HB, red clump (RC)) can also influence the accuracy. Detailed discussions of computing random and systematic uncertainties for CMD-based SFHs are described in \citet{dolphin2013} and \citet{dolphin2012}, respectively.  Here we provide a brief description of how we computed uncertainties for this sample.

Random uncertainties were computed using a hybrid Monte Carlo (HMC) process \citep{duane1987}, with implementation details as described by \citet{dolphin2013}.  The result of this Markov chain Monte Carlo routine is a sample of 10$^4$ SFHs whose density is proportional to the probability density, i.e., the density of samples is highest near the maximum likelihood point.  Upper and lower random error bars for any given value (e.g., cumulative star formation at a particular point in time) are calculated by identifying the boundaries of the highest-density region containing 68\% of the samples, where the value 68\% adopted is an equivalent percentile range to the $\pm$ 1-$\sigma$ bounds of a normal distribution. 

Systematic uncertainties were computed by running 50 Monte Carlo SFH realizations on the best fit model CMD.  For each realization, the model CMD was shifted in $\log(\mathrm{T_{eff}})$ and $\mathrm{M_{bol}}$ following  the procedure detailed in \citet{dolphin2012} using the values listed in Table \ref{tab2}, and the SFH was re-measured.  From the resulting 50 SFHs, we computed the range that contained 68\% of distributions, and designated this as the systematic uncertainties.  As discussed in \citet{dolphin2012}, this technique is designed to produce 1-$\sigma$ uncertainties that include the best fit SFHs as derived with either Padova, BaSTI \citep{pietrinferni2004}, or Dartmouth stellar evolution models.

For clarity, throughout the paper, we provide values for both the systematic and random uncertainties, and will specifically refer to the type of uncertainty being considered in the context of particular discussions.   The values of these uncertainties are tabulated in Tables \ref{tab:field_sfhs} and \ref{tab:whole_sfhs}.

Finally, we also estimate systematic uncertainties for various ensemble populations, e.g., the average cumulative SFH of dIrrs, that we analyze throughout this paper.  Systematic uncertainties for ensemble SFHs are challenging to compute.  Simply adding systematic uncertainties in quadrature from individual SFHs does not account for covariances in the uncertainties that can arise for galaxies with similar photometric depths and/or SFHs, and would result in an overestimate of the true uncertainty.  

To estimate the systematics for the ensemble populations, we adopted the following Monte Carlo approach which was first outlined in \citet{weisz2011b}.  Using the BaSTI stellar models, we generate 50 CMDs, each with a constant SFH and 100,000 stars, to minimize Poisson noise.  We then recover the simulated SFH using the Padova models.  We repeat this exercise at a variety of photometric depths to capture the systematic uncertainties as a function of age and photometric depth.  We compute the mean and standard deviation for each realization.

For each ensemble population (e.g., the photometric depth distribution of all the real dIrrs), we then compute the ensemble mean and standard deviation, i.e., we compute the average offset and standard deviation at each photometric depth.  Here, the standard deviation is the root mean squared of the standard deviation for a set of Monte Carlo realizations of a given depth.

Using these ensemble uncertainties, we then compute two useful systematic uncertainties, differential (or relative) and absolute.  The differential systematic is used when comparing two different populations, e.g., dSph and dIrr SFHs.  In this case, we subtract the input constant SFH from the mean of the ensemble SFHs.  The differential systematic, i.e., relative to a known truth, is then the root square sum (RSS) of the difference in the mean and ensemble standard error.  The second is the absolute systematic, i.e., not relative to a known true SFH, which enables comparisons with non-CMD based SFH determinations.   This is computed identically to the differential systematic, only the true mean is not subtracted out.  The absolute systematic is then the RSS of the recovered mean and the standard deviation.  

We provide a simple example to illustrate the combination of the ensemble uncertainties. Suppose we have three groups of galaxies: X, Y, and Z, each with two galaxies, and we wish to compute the systematic of group X relative to groups Y and Z.  In group X, there are two galaxies.  At as single time point, 10 Gyr ago, the first has a mean of $-$0.1, relative to the input constant SFH and a standard deviation of 0.1. At the same epoch, the second has a mean of $+$0.1 and a standard deviation of 0.2.  The average of the two has a mean of 0 and a standard deviation of 0.22.  Now we wish to know the relative uncertainty of group X relative to Y and Z.  Suppose groups Y and Z have means of $-$0.05 and $-$0.10, so the combined mean is $-$0.05.  Relative to this average, group X has a mean of $+$ 0.05 and a standard deviation of 0.22.  The total differential systematic at 10 Gyr has a values of 0.23, which is the RSS of offset (0.05) and standard deviation (0.22).

\begin{figure}
\begin{center}
\plotone{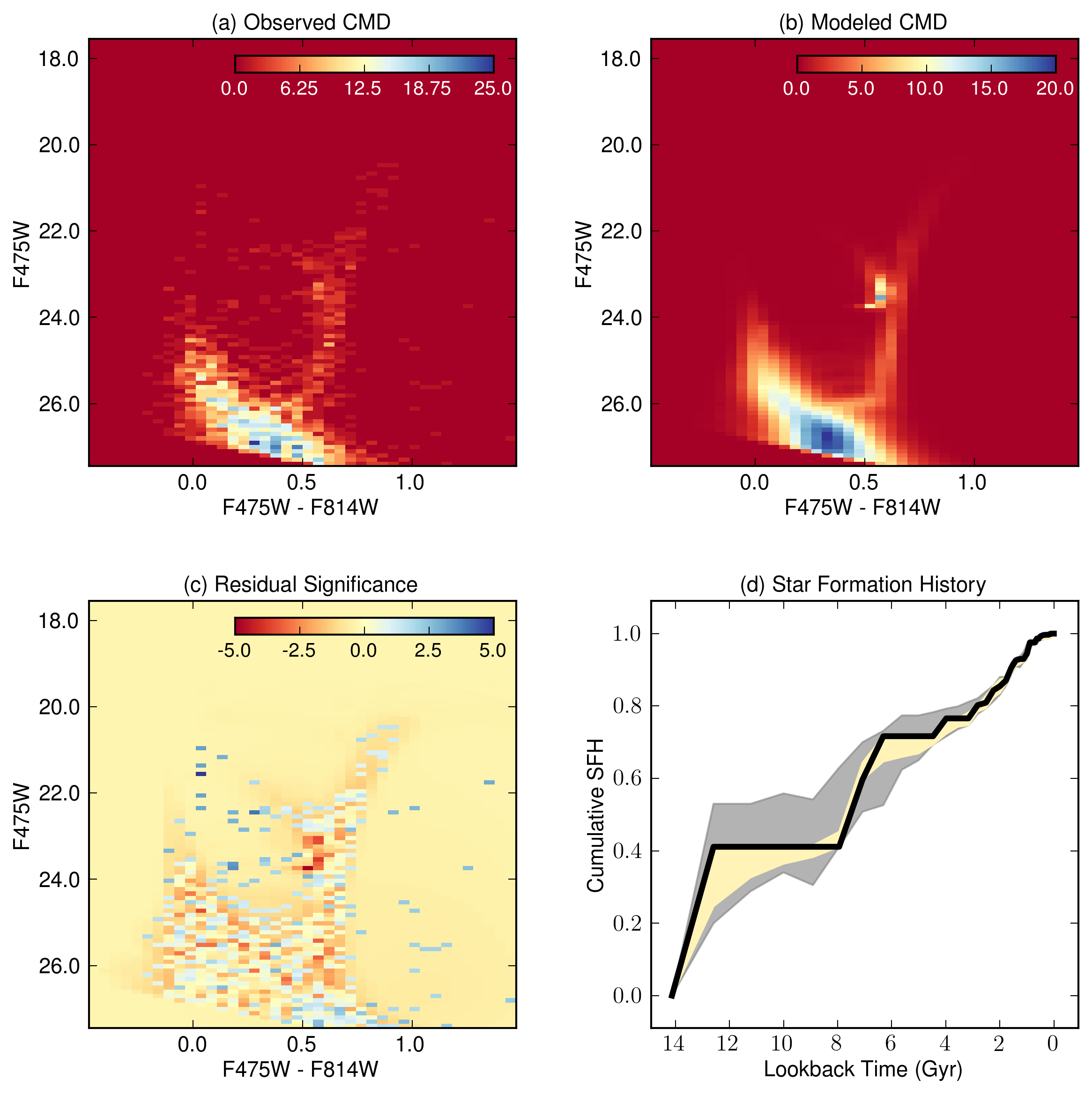}
\caption{\small{An illustration of the SFH fitting process using Leo~T.  Panel (a): the observed Hess diagram; Panel (b) Best fit model Hess Diagram; Panel (c) Residual significance Hess diagram; Panel (d) The cumulative SFH, i.e., fraction of the total stellar mass formed prior to a given epoch.  In Panels (a) and (b) the color-scale reflects the number of stars. In Panel (c) the scaling reflects the significance of each pixel in the residual relative to the standard deviation of a Poisson distribution. In Panel (d) the best fit SFH is the solid black line and the error envelopes represent the 68\% confidence interval around the best fit due to random (yellow) and total (random and systematic; grey) uncertainties.  Overall, the best fit model does an excellent job of reproducing the observed Hess diagram of Leo~T, as indicated by the lack of strong coherent structure in the Panel (c). The process of SFH measurement and uncertainty analysis is discussed in \S \ref{sec:analysis}.}}
\label{fig:leot_hess}
\end{center}
\end{figure}

\subsection{An Example Star Formation History Measurement}
\label{sec:sfhexample}

In Figure \ref{fig:leot_hess}, we show an example of the quality of our CMD fit for Leo~T.  Leo~T contains several elements that are challenging to the CMD fitting process, including a sparsely populated CMD and star formation at virtually all ages.  It therefore provides a good illustration of how CMD fit qualities are evaluated.  In this figure, we show Hess diagrams (i.e., a 2D density plot of the CMD) for: the observed data in panel (a); the best fit model in panel (b); and the residual significance (i.e., the residual weighted by the variance) in panel (c).  We also plot the resulting cumulative SFH in panel (d).  

Overall, we find that the model reproduces the observed CMD reasonably well.  The best fit model contains an RGB, SGB, and oldest MSTO that are well-matched to the observations in terms of density of stars, luminosities/colors, and width of features.  For intermediate age populations, the model shows several turnoffs crossing horizontally from the MS to the RGB, that match the scattering of observed stars over the same range.  Finally, the model finds a sparsely populated blue helium burning sequence (BHeB), which is in reasonable agreement with the data, but that is subject to the known mismatches in color between BHeB models and observations \citep[e.g.,][]{mcquinn2011}.  Even with this complex CMD, the observed and model Hess diagrams agree well, with few coherent structures in the residual Hess diagram.

However, not all features are in good agreement. As shown in the residual Hess diagram, the presence of coherent structure near the RC indicates a systematic mismatch between the observed and modeled RC.  This problem is a known feature of the generation of Padova models used in this paper \citep[e.g.,][]{gallart2005}.  Similarly, the Padova RGB is often too blue compared to observations, which can result in a mismatch between photometric and spectroscopic metallicities, as discussed, for example, in \citet{weisz2012b}.  In the case of Leo~T, the model RGB matches the data reasonably well, but does so with a higher mean metallicity ([M/H] $=$ $-$1.7) than the spectroscopic mean metallicity of [Fe/H] $\sim$ $-$2.3 \citep[e.g.,][]{simon2007}.   Both issues may be alleviated by newer models \citep[e.g.,][]{bressan2012, vandenberg2012, paxton2013}, but this hypothesis remains largely untested at the present.  However, aside from the Padova models with low metallicity AGB updates in \citet{girardi2010}, there are no other stellar evolution libraries that include the entire range of mass, metallicity, and evolutionary phases needed to analyze a diverse collection of galaxies such as the LG dwarfs.

Leo~T's cumulative SFH, i.e., the fraction of total stellar mass formed prior to a given epoch, is shown in panel (d).  The black line represents the best fit SFH and the error envelopes reflect the 68\% confidence interval of random uncertainties (yellow) and the total uncertainties (random plus systematic; grey), as described in \S \ref{sec:sfherrors}. The cumulative SFHs provide a convenient way to compare galaxies whose absolute SFHs and stellar masses are orders of magnitude apart.  Cumulative SFHs also minimize many of the issues that affect interpreting absolute SFHs, e.g., interpreting covariant SFRs in adjacent time bins and defining an appropriate time resolution \citep[e.g.,][]{harris2001, dolphin2002, mcquinn2010b}.  

We inspected the fits for every field using the same diagnostics as in Figure \ref{fig:leot_hess}.  Overall, we found the fits to be satisfactory, with similar caveats to those listed above (e.g., systematic mismatches in the RC and offsets in the helium burning sequences in star-forming galaxies, and higher mean metallicities than those derived from spectroscopy).

\subsection{The Anomalous Star Formation Histories of Cetus and And {\sc VII}}
\label{sec:tpagb}

Despite the satisfactory quality of the fits, two of our galaxies (And {\sc VII} and Cetus) appeared to have unrealistically precise systematic uncertainties given their extremely shallow CMDs.  In this section, we discuss the causes of the overly precise uncertainties and discuss the implications for our ensemble of SFHs. 

The CMDs of these two galaxies are shallow enough to exclude age sensitive features such as the HB or RC. Additionally, as early type galaxies, they have no young stellar sequences. Therefore, measurement of their SFHs relied exclusively on the RGB and Thermally Pulsating-AGB (TP-AGB) populations.  We have found that the Padua isochrones downloaded from the CMD web tool\footnote{\textcolor{blue}{\url{http://stev.oapd.inaf.it/cgi-bin/cmd}}} included as few as two TP-AGB points brighter than the TRGB, introducing artifacts above the TRGB into the synthetic CMD models. These artifacts were minimized for the oldest populations, causing \texttt{MATCH} to prefer to use the oldest populations when fitting these CMDs.  Because And {\sc VII} and Cetus are the only early type galaxies in our sample that have such shallow CMDs, their SFHs were therefore measured to be dominated by ancient stars to an unrealistically high precision.

\begin{figure}
\begin{center}
\plotone{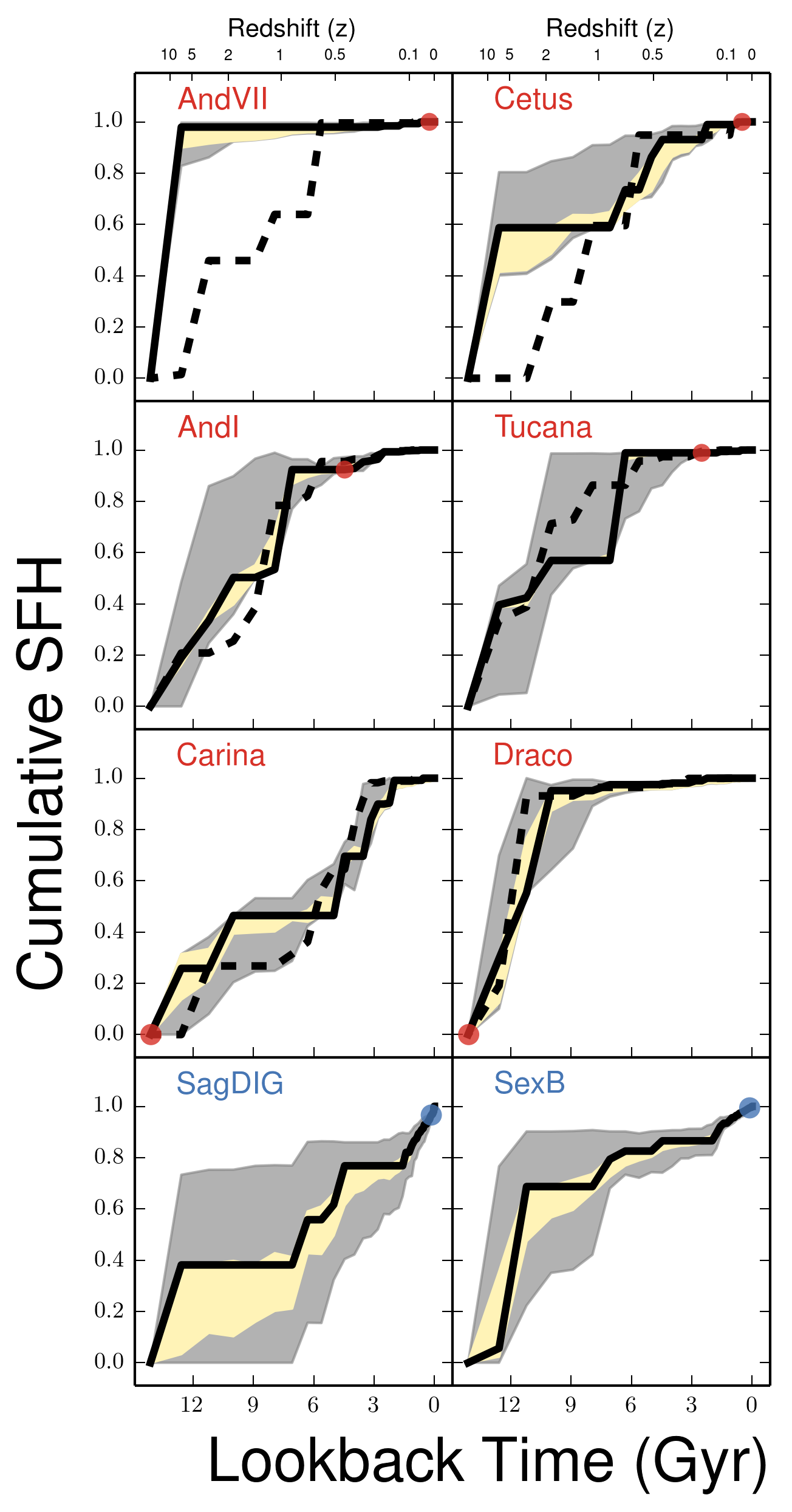}
\caption{\scriptsize{The effects of TP-AGB models on SFH measurements.  The solid black lines are SFHs measured with the Padova models and the dashed lines are measured with the Dartmouth models.  The yellow and grey envelopes reflect the 16th and 84th confidence intervals of the random and total (random plus systematic) uncertainties in the Padova SFHs.  The colored dots in each panel indicate the approximate age of the oldest MSTO on each CMD, and are color-coded by morphological type (dSph, red; dIrr, blue). The SFHs of And {\sc VII} and Cetus have overly precise uncertainties given their extremely shallow CMDs due to artifacts in the TG-AGB models.  A comparison with the Dartmouth solutions demonstrates what the true amplitude of the uncertainties should be.  The TP-AGB artifacts only affect CMDs of early type galaxies where the age leverage comes entirely from the RGB and TP-AGB.  Galaxies with moderately deep CMDs (e.g., And {\sc II}, Tucana) or extremely deep CMDs (e.g., Carina, Draco) do not suffer from this problem.  Likewise, star-forming galaxies with extremely shallow CMDs (e.g., Sag~DIG, Sex~B) show appropriately sized uncertainties.  In our sample, only And {\sc VII} and Cetus are affected by this issue.  The remaining 95\% have appropriately sized uncertainties given their CMD depths.  See \S \ref{sec:tpagb} for further discussion.}}
\label{fig:tpagb}
\end{center}
\end{figure}

To illustrate our concerns with these galaxies, we plot their SFHs in the top two panels Figure \ref{fig:tpagb}. For comparison, we have also measured their SFHs using the Dartmouth models (dashed lines).  Here, we see that the Padova and Dartmouth SFHs of And {\sc VII} and Cetus are not consistent within uncertainties, because the uncertainties are significantly underestimated for reasons described above.

In 95\% of our sample, the uncertainties have the appropriate amplitude.  We illustrate this for representative CMDs in the bottom panels of Figure \ref{fig:tpagb}.  The middle panels show the case of early type galaxies with intermediate depth CMDs (i.e., they include the HB; e.g., And {\sc I} and Tucana) and extremely deep CMDs (i.e., they include the oldest MSTO; e.g., Carina and Draco).  In all cases, the SFHs are consistent within uncertainties.  In the bottom panels, we highlight that shallow CMDs of star-forming galaxies (Sex~B and Sag~DIG) also have the appropriate amplitude uncertainties.  Thus the effects of TP-AGB artifacts on SFH measurements are limited to a minority of cases, but general caution is suggested when analyzing similarly shallow CMDs of early type systems, which echoes previous discussion by \citet{gallart2005}.  For the remainder of this paper, we will exclude the SFHs of Cetus and And {\sc VII} from any analysis.

\subsection{Comparisons with Star Formation Histories Derived from Deeper CMDs}
\label{sec:lcidcompare}

One way to illustrate the robustness of our SFHs is to compare those derived from typical depths to SFHs derived from CMDs that extend below the oldest MSTO.  This test is particularly useful in cases where the WFPC2 CMDs do not reach the oldest MSTO and/or are sparsely populated, each of which can introduce large random and/or systematic uncertainties. Fortunately, within the past few years there have been several HST/ACS observations of LG dwarfs in our sample that extend below the oldest MSTO, enabling such an empirical exercise. 

\begin{figure}
\begin{center}
\plotone{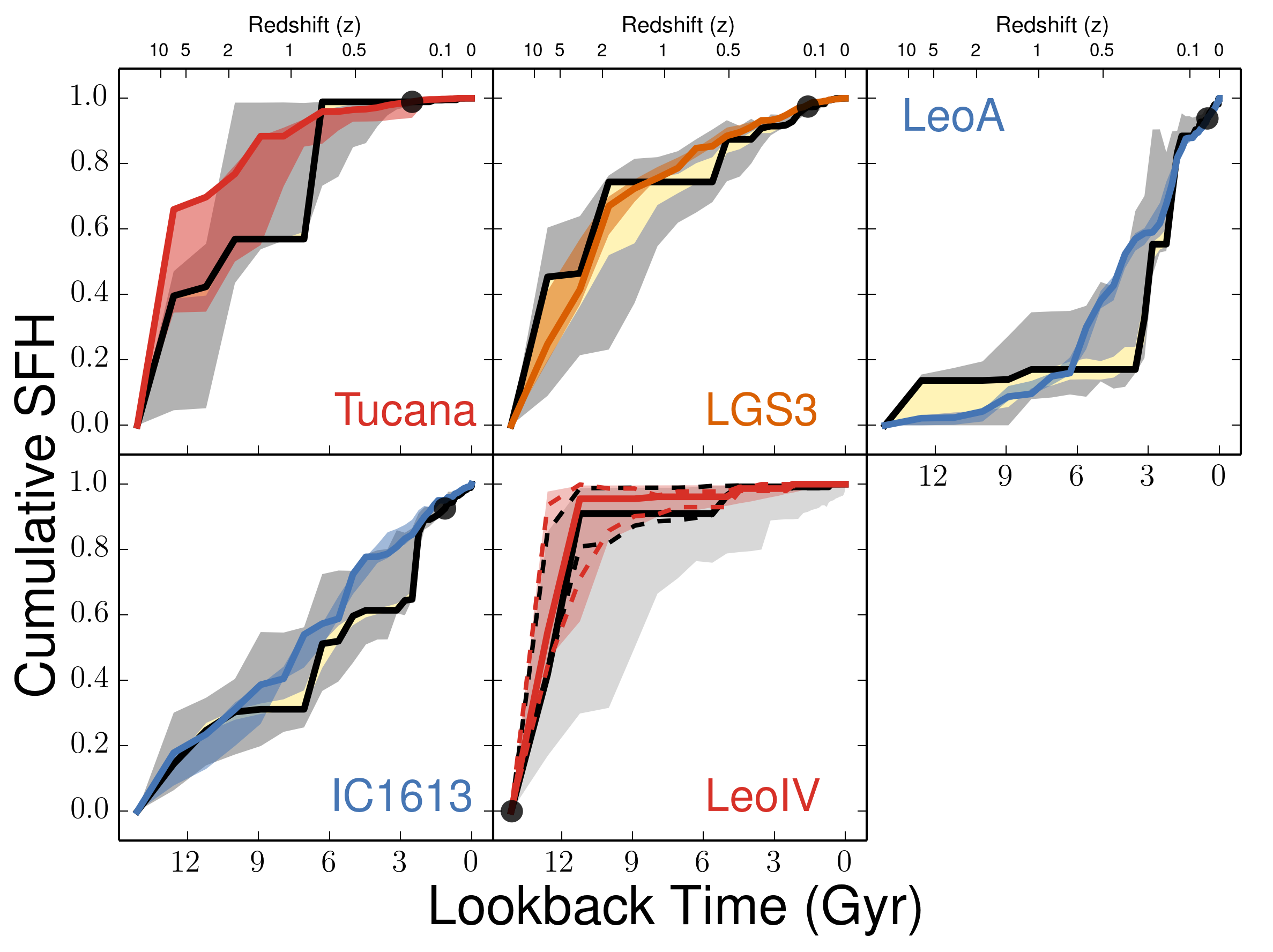}
\caption{\small{A comparison between the SFHs analyzed in this paper (black and grey) with SFHs derived from deeper HST/ACS observations (colored; dSphs, red; dIrrs, blue; dTrans, orange) of the same galaxies.  For the WFPC2 data, the uncertainties are as described in Figure \ref{fig:tpagb}.  For the deeper data, the shaded error envelopes indicate the total uncertainties (i.e., random and systematic). The black dot in each panel indicates the approximate MSTO age for the WFPC2 data.  Overall, each pair of WFPC2 and ACS SFHs are consistent within uncertainties, indicating good agreement.  In the case of Leo {\sc IV}, both datasets extend below the oldest MSTO by $\sim$ 1 mag for WFPC2 and 3-4 mag for ACS.  Despite the depth difference, the amplitude of the systematic uncertainties (dashed lines) are identical, illustrating the stars below the MSTO do not help improve the SFH measurement.  Instead, the improved precision in the ACS SFH is due to the increased number of stars on the CMD around the MSTO and SGB.}}
\label{fig:comp_deep}
\end{center}
\end{figure}

In Figure \ref{fig:comp_deep}, we consider five such cases. Here, we plot the cumulative SFHs measured from our data in greyscale and SFHs derived from deeper data in color.  For Tucana, LGS~3, Cetus, Leo~A, and IC~1613, the deeper observations come from the LCID program \citep[e.g.,][]{cole2007, monelli2010b,monelli2010c, hidalgo2011, skillman2014} and for Leo {\sc IV} the observations are from \citet{brown2012}.  In all cases, the CMDs extend one or more magnitudes below the oldest MSTO.  To ensure consistency, reduction of the ACS data was done with \texttt{DOLPHOT}, the successor to \texttt{HSTPHOT} that has been extended to include a specific ACS module, and the SFHs were measured using the identical process described in \S \ref{sec:measuresfhs}.  All error envelopes on the deeper data in Figure \ref{fig:comp_deep} reflect the total uncertainties in the SFHs (i.e., systematic and random combined).  

Overall, we find that the WFPC2-based SFHs are similar to those derived from the deeper data.  The general trends present in the deep ACS-based SFHs are reflected in the shallower data.   For example, Leo~A shows a late time increase in both SFHs, despite a WFPC2 CMD that only includes a $\sim$ 1 Gyr MSTO.   Additionally, IC~1613, LGS~3, Tucana, and Leo {\sc IV} all have WFPC2- and ACS-based SFHs that are in excellent agreement, i.e., all are consistent within the uncertainties. 

\begin{figure*}[th!]
\begin{center}
\plotone{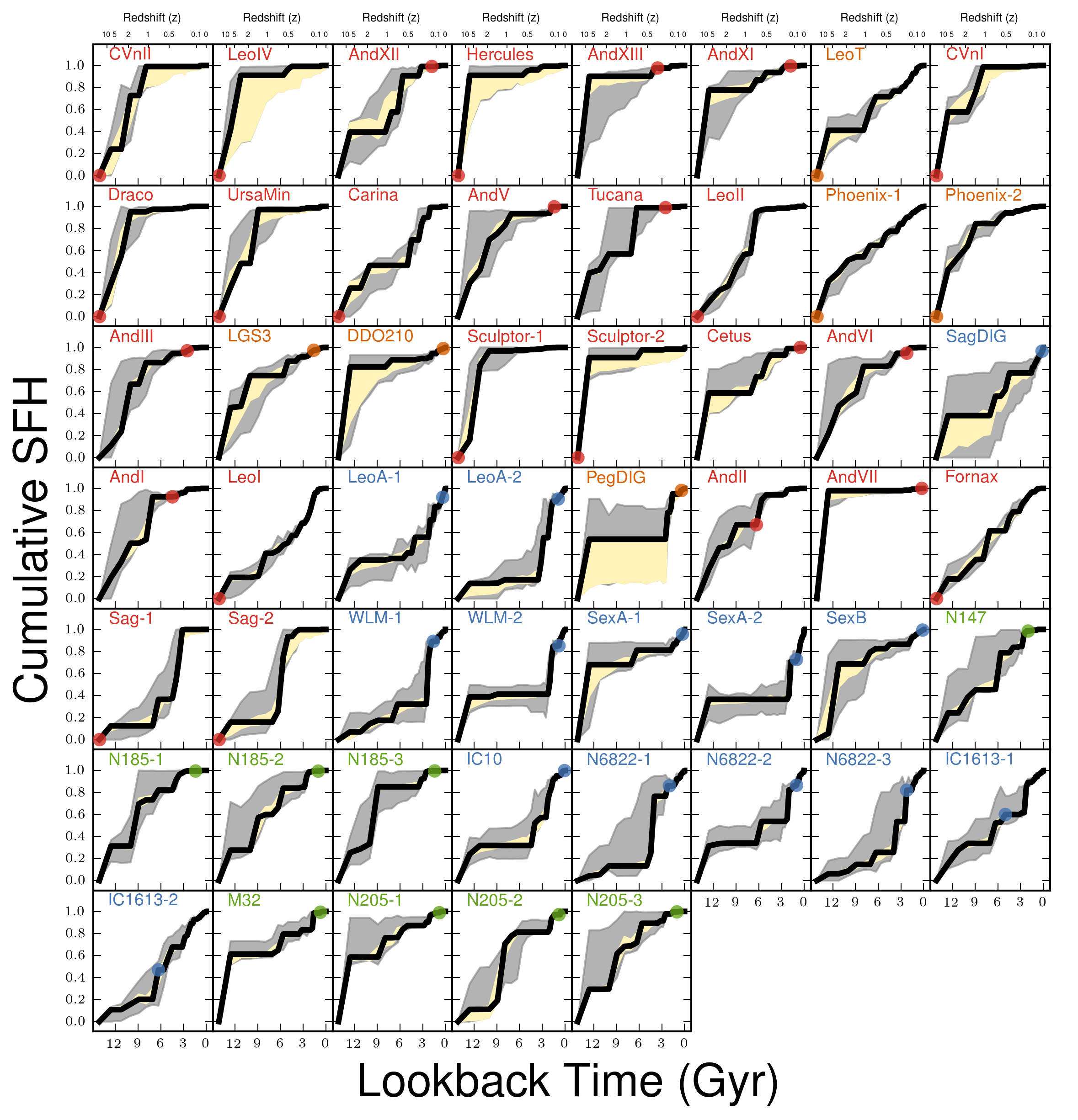}
\caption{\small{The cumulative SFHs of all fields analyzed in this paper, sorted by increasing absolute V-band luminosity, assuming the distance and extinction corrections listed in Table \ref{tab2}.  The fields are color-coded according to present day morphological type: dSph (red); dIrr (blue); dTrans (orange); dE (green). The solid black line represents the best fit SFH, while shaded envelops represent the 68\% confidence interval due to random uncertainties (yellow) and total uncertainties (random and systematic; grey). Each plot has been normalized such that its total stellar mass formed is unity at the present day.  The colored dots in each panel indicate the approximate age of the oldest MSTO on each CMD. The tabulated SFHs and uncertainties for each field are listed in Table \ref{tab:field_sfhs}.}}
\label{fig:sfhgrid}
\end{center}
\end{figure*}

Comparing the SFHs of these five galaxies provides two important insights into CMD analysis.  First, CMDs that extend only as deep as the HB (i.e., they include the RC and HB), and do not include the oldest MSTO, still contain significant information about the lifetime SFH of a galaxy.  That is, SFHs based on CMDs that only contain less certain phases of evolution (e.g., the RGB, AGB) along with the HB and/or RC, result in an accurate but less precise measurement.  Second, including systematic uncertainties is essential for accurate errors, given that they typically dominante the total error budget.  Had we not included them in our analysis, the uncertainties on the shallow CMDs would have been significantly underestimated, and the SFHs from the shallower CMDs would have appeared to be inconsistent with the SFHs derived from deeper data.  We refer the reader to to several other papers for in-depth discussions of systematic uncertainties \citep[e.g.,][]{aparicio2009, weisz2011a, dolphin2012}.

\begin{figure*}
\begin{center}
\plotone{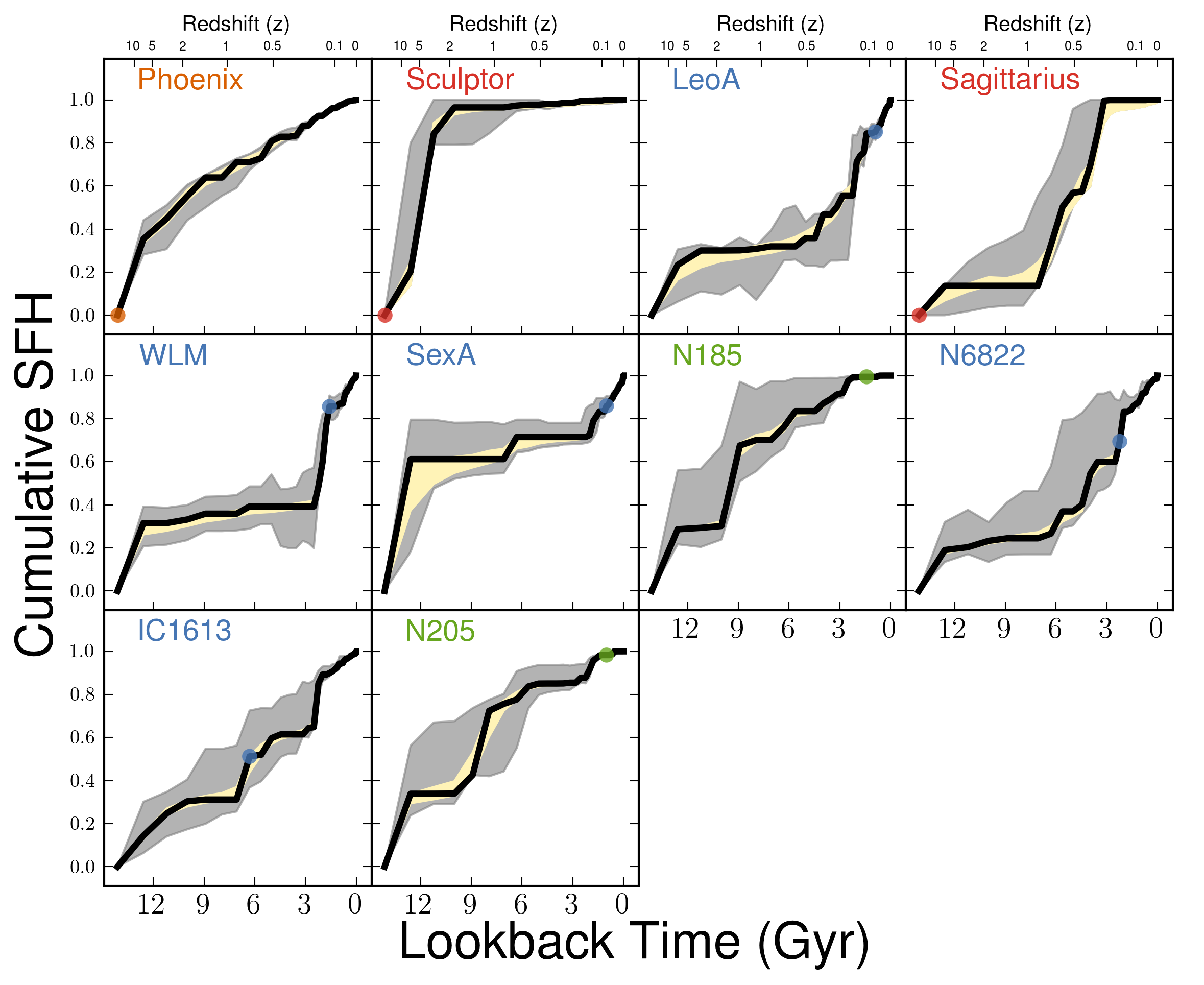}
\caption{The same as Figure \ref{fig:sfhgrid}, only for the combined SFHs of galaxies with multiple fields.  The colored dots in each panel indicate the approximate age of the oldest MSTO available for each galaxy. The tabulated SFHs and uncertainties for each galaxy are listed in Table \ref{tab:whole_sfhs}.}
\label{fig:galaxysfhgrid}
\end{center}
\end{figure*}

In the final panel of Figure \ref{fig:comp_deep}, we compare the SFHs of Leo {\sc IV} derived from two datasets that include the oldest MSTO and overlap in their sky coverage.  The differences in the datasets are two-fold.  First, the ACS CMD extends several magnitudes below the oldest MSTO, whereas the WFPC2 CMD extends $\sim$ 1 mag below the oldest MSTO.  Second, the ACS CMD contains more stars, due to the larger ACS field of view.

Despite these differences, the SFHs from the two datasets are almost identical.  Both best fit SFHs show that Leo {\sc IV} formed 90\% of its stellar mass prior to $\sim$ 11-12 Gyr ago.  The primary difference between the two solutions is in the amplitudes of the uncertainties.  For both SFHs, we have plotted the total uncertainties as shaded envelopes (ACS in red; WFPC2 in grey) and the systematic-only uncertainties as dashed lines (ACS in red; WFPC2 in black).  The total uncertainties for the WFPC2 solution are much larger than for the ACS solution.  However, the systematic uncertainties are virtually identical in amplitude, since they are typically dominated by uncertainties in the model isochrones which affect both datasets identically.  Thus, the improved precision in the ACS-based SFH is due to the increased number of stars on the CMD, and not to the extra depth gained.  Stars below the oldest MSTO are largely insensitive to SFH \citep[e.g.,][]{gallart2005}, and therefore do not provide additional constraints on the SFH of a galaxy.  In contrast, the increased number of stars in age sensitive regions of the CMD (e.g., RGB, SGB, MSTO) can drastically decrease the random uncertainties, particularly in the case of sparsely populated ultra-faint dwarf galaxies.  For the purposes of measuring globally representative SFHs, it is preferable to survey larger areas that just capture the oldest MSTO than it is to acquire extremely deep observations extending well below the oldest MSTO of a smaller area.

\section{Characterizing the Star Formation Histories}
\label{sec:results}

In Figure \ref{fig:sfhgrid} we plot the cumulative SFHs for all fields in our sample.  In Figure  \ref{fig:galaxysfhgrid}, we plot combined cumulative SFHs for galaxies that have multiple fields.  The combined SFHs are simply the unweighted sum of the absolute SFHs of each field, converted into cumulative space. The uncertainties were propagated following the method described in the appendix of \citet{weisz2011a}.  Figures \ref{fig:sfhgrid} and \ref{fig:galaxysfhgrid} show that the LG dwarf galaxies have a wide variety of complex SFHs, which is consistent with previous studies of nearby dwarf galaxies \citep[e.g.,][]{hodge1989, grebel1997, mateo1998, tosi2007, tolstoy2009, mcquinn2010, weisz2011a}.  

In the following sections, we explore the complexity of these SFHs in the context of several intrinsic (e.g., stellar mass) and external (e.g., proximity to a massive galaxy) variables. Our primary focus is to investigate trends in the ensemble SFHs as opposed to individual galaxies.  Many papers have been devoted to studies of individual LG dwarf SFHs, as indicated in Table \ref{tab1}.  While comparing our measurements with those in past studies is an interesting exercise, it is beyond the scope of the present paper.  Instead, we focus on the broader properties of the sample, and only comment on the SFHs of individual galaxies in specific instances.  To facilitate use of this dataset for comparison with existing SFHs, models of dwarf galaxies, and/or LG simulations, we tabulate the detailed SFHs for each field and galaxy, including all uncertainties, in Tables \ref{tab:field_sfhs} and \ref{tab:whole_sfhs}.

\begin{figure*}
\begin{center}
\plotone{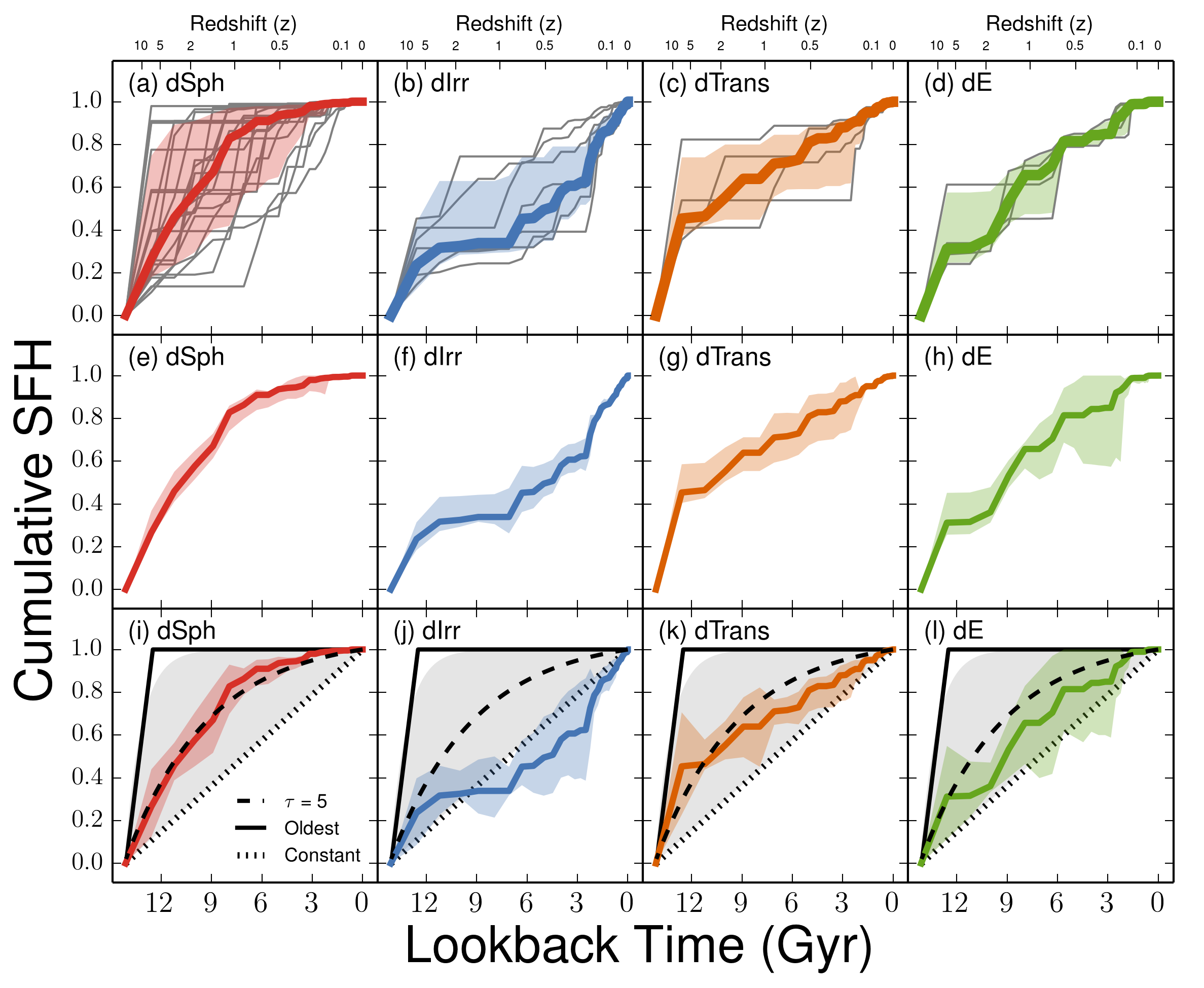}
\caption{The unweighted average (i.e., number weighted) SFHs per morphological type (dSph, red; dIrr, blue; dTrans, orange; dE, green).  The solid lines represent the median value of the best fit SFHs from individual galaxies.  Panels (a) - (d): The uncertainties are the 16th and 84th percentiles of the distribution of best fit SFHs, i.e., statistical uncertainties.  The grey lines indicate the individual SFHs in each morphological group.  Interestingly, dSphs show the largest scatter, suggesting that the formation of dSphs may be a complex process.  Panels (e) - (h): Uncertainties reflect the root square sum (RSS) of the differential ensemble systematics and the standard error in the median.  This allows for a relative comparison between groups.  Panels (i) - (l): The average SFHs with simple SFH models over plotted: purely ancient (solid black), lifetime constant (dashed), exponentially declining `$\tau$-model' (dotted, $\tau =$ 5 Gyr; grey, $\tau =$ 0.1 - 100 Gyr).  Uncertainties are the RSS of the absolute ensemble systematics and the standard error in the median. The dSphs are well-described by a simple model with $\tau =$ 5 Gyr, while the other types require a mix of a $\tau$ model $\gtrsim$ 12 Gyr ago ($\tau$ $\sim$ 3 Gyr) with an approximately constant SFH subsequently.}
\label{fig:avg_sfhs}
\end{center}
\end{figure*}

\begin{deluxetable*}{ccccccccc}
\tablecaption{Observational Properties of each Sample Galaxy}
\tablehead{
\colhead{Galaxy} &
\colhead{Morphological} &
\colhead{(m-M)$_0$} &
\colhead{A$_V$} &
\colhead{M$_{\mathrm{V}}$} &
\colhead{M$_{\star}$} &
\colhead{M$_{\mathrm{H{\sc I}}}$} &
\colhead{Distance to Nearest Host} &
\colhead{r$_{\mathrm{h}}$}  \\
\colhead{Name} &
\colhead{Type} &
\colhead{} &
\colhead{(mag)} &
\colhead{} &
\colhead{(10$^{6}$ \msun)} &
\colhead{(10$^{6}$ \msun)} &
\colhead{(kpc)} &
\colhead{(pc)}  \\
\colhead{(1)} &
\colhead{(2)} &
\colhead{(3)} &
\colhead{(4)} &
\colhead{(5)} &
\colhead{(6)} &
\colhead{(7)} &
\colhead{(8)} &
\colhead{(9)} 
}
\startdata 
Andromeda {\sc I} & dSph & 24.36$\pm$0.07 & 0.15 & -11.7 & 3.9 & 0.0 & 58 & 672 \\
Andromeda {\sc II} & dSph & 24.07$\pm$0.06 & 0.17 & -12.4 & 7.6 & 0.0 & 184 & 1176 \\
Andromeda {\sc III} & dSph & 24.37$\pm$0.07 & 0.15 & -10.0 & 0.83 & 0.0 & 75 & 479 \\
Andromeda {\sc V} & dSph & 24.44$\pm$0.08 & 0.34 &  -9.1 & 0.39 & 0.0 & 110 & 315 \\
Andromeda {\sc VI} & dSph & 24.47$\pm$0.07 & 0.17 &  -11.3 & 2.8 & 0.0 & 269 & 524 \\
Andromeda {\sc VII}  & dSph & 24.41$\pm$0.10 & 0.53 & -12.6 & 9.5 & 0.0 & 218 & 776 \\
Andromeda {\sc XI} & dSph & 24.40$_{-0.5}^{+0.2}$ & 0.21 &  -6.9 & 0.049 & 0.0 & 104 & 157 \\
Andromeda {\sc XII} & dSph & 24.47$\pm$0.30 & 0.30 &  -6.4 & 0.031 & 0.0 & 133 & 304 \\
Andromeda {\sc XIII} & dSph & 24.80$_{-0.4}^{+0.1}$ & 0.23 & -6.7 & 0.041 & 0.0 & 180 & 207 \\
Carina & dSph & 20.11$\pm$0.13 & 0.17 & -9.1 & 0.38 & 0.0 & 107 & 250 \\
Cetus & dSph & 24.38$\pm$0.07 & 0.08 & -11.2 & 2.6 & 0.0 & 756 & 703 \\
Canes~Venatici {\sc I} & dSph & 21.69$\pm$0.10 & 0.04 &  -8.6 & 0.23 & 0.0 & 218 & 564 \\
Canes~Venatici {\sc II} & dSph & 21.02$\pm$ 0.06 & 0.03 & -4.9 & 0.0079 & 0.0 & 161 & 74 \\
DDO~210 & dTrans & 25.15$\pm$0.08 & 0.14 & -10.6 & 1.6 & 4.1 & 1066 & 458 \\
Draco & dSph & 19.40$\pm$0.17 & 0.07 & -8.8 & 0.29 & 0.0 & 76 & 221 \\
Fornax & dSph & 20.84$\pm$0.18 & 0.06 & -13.4 & 20.0 & 0.0 & 149 & 710 \\
Hercules & dSph & 20.60$\pm$0.20 & 0.17 & -6.6 & 0.037 & 0.0 & 126 & 330 \\
IC~10 & dIrr & 24.50$\pm$0.12 & 4.3 & -15.0 & 86.0 & 50.0 & 252 & 612 \\
IC~1613 & dIrr & 24.39$\pm$0.12 & 0.07 & -15.2 & 100.0 & 65.0 & 520 & 1496 \\
Leo~A & dIrr & 24.51$\pm$0.12 & 0.06 & -12.1 & 6.0 & 11.0 & 803 & 499 \\
Leo~{\sc I} & dSph & 22.02$\pm$0.13 & 0.10 & -12.0 & 5.5 & 0.0 & 258 & 251 \\
Leo~{\sc II} & dSph & 21.84$\pm$0.13 & 0.05 & -9.8 & 0.74 & 0.0 & 236 & 176 \\
Leo~{\sc IV} & dSph & 20.94$\pm$0.09 & 0.07  & -5.8 & 0.019 & 0.0 & 155 & 206 \\
Leo~T & dTrans & 23.10$\pm$0.10 & 0.09 &  -8.0 & 0.14 & 0.28 & 422 & 120 \\
LGS~3 & dTrans & 24.42$\pm$0.07 & 0.11 & -10.1 & 0.96 & 0.38 & 269 & 470 \\
M32 & dE & 24.53$\pm$0.21 & 0.17 & -16.4 & 320.0 & 0.0 & 23 & 110 \\
NGC~147 & dE & 24.15$\pm$0.09 & 0.47 & -14.6 & 62.0 & 0.0 & 142 & 623 \\
NGC~185 & dE & 23.95$\pm$0.09 & 0.50 & -14.8 & 68.0 & 0.11 & 187 & 42 \\
NGC~205 & dE & 24.58$\pm$0.07 & 0.17 &-16.5 & 330.0 & 0.4 & 42 & 590 \\
NGC~6822 & dIrr & 23.31$\pm$0.08 & 0.65 & -15.2 & 100.0 & 130.0 & 452 & 354 \\
Pegasus & dTrans & 24.82$\pm$0.07 & 0.19 & -12.2 & 6.61 & 5.9 & 474 & 562 \\
Phoenix & dTrans & 23.09$\pm$0.10 & 0.04 & -9.9 & 0.77 & 0.12 & 415 & 454 \\
Sag~DIG & dIrr & 25.14$\pm$0.18 & 0.34 & -11.5 & 3.5 & 8.8 & 1059 & 282 \\
Sagittarius & dSph & 17.10$\pm$0.15 & 0.42 & -13.5 & 21.0 & 0.0 & 18 & 2587 \\
Sculptor & dSph & 19.67$\pm$0.14 & 0.05 & -11.1 & 2.3 & 0.0 & 89 & 283 \\
Sex~A & dIrr & 25.78$\pm$0.08 & 0.12 & -14.3 & 44.0 & 77.0 & 1435 & 1029 \\
Sex~B & dIrr & 25.77$\pm$0.03 & 0.09 &  -14.4 & 52.0 & 51.0 & 1430 & 440 \\
Tucana & dSph & 24.74$\pm$0.12 & 0.09 & -9.5 & 0.56 & 0.0 & 882 & 284 \\
Ursa~Minor & dSph & 19.40$\pm$0.10 & 0.09 & -8.8 & 0.29 & 0.0 & 78 & 181 \\
WLM & dIrr & 24.85$\pm$0.08 & 0.10 & -14.2 & 43.0 & 61.0 & 836 & 2111
\enddata
\tablecomments{Physical properties for galaxies in our sample taken from \citet{mcconnachie2012}.  The stellar mass listed in column (4) is computed from the integrated V-band luminosity and assumes \msun/\lsun $=$ 1. } 
\label{tab:alan_data}
\end{deluxetable*}

\subsection{The Average Star Formation Histories}
\label{sec:averagesfhs}

We begin by examining average SFHs for different subsets of galaxies.  Average SFHs provide a simple diagnostic for broadly differentiating between galaxies of different morphological types, for comparing to simple models of star formation, and for tracing star formation as a function of halo mass over cosmic time \citep[e.g.,][]{grebel2003, weisz2011a, weisz2011b, leitner2012, behroozi2013}.  

In Figure \ref{fig:avg_sfhs}, we plot the unweighted average cumulative SFHs of galaxies in our sample, grouped by morphological type; uncertainties are the 68\% dispersion in the best fit SFHs.  We first consider the relationships between the average SFH and scatter from individual galaxies per morphological type in Panels (a) - (d). 

This comparison reveals several interesting trends. First, dSphs are predominantly old systems and have formed the vast majority of their stars prior to z $\sim$ 2 (10 Gyr ago).  However, while this behavior holds on average, the individual galaxies show significant scatter, ranging from purely old to those with nearly constant lifetime SFHs.  The rich variety in dSph SFHs has been known for over a decade \citep[e.g.,][]{grebel1997, mateo1998, grebel2003}, and has motivated increasingly sophisticated model efforts to reproduce the properties of MW dSphs \citep[e.g.,][]{mayer2001, mayer2006, zolotov2012, brooks2013, kazantzidis2013}.

Second, there are few predominantly old dIrrs.  On average, dIrrs formed $\sim$30\% of their stellar mass prior to z $\sim$ 2, and show an increasing SFR toward the present, beginning around z $\sim$ 1 (7.6 Gyr ago).  This behavior is generally reflected in the SFHs of individual dIrrs, which show only modest scatter relative to the average. The small degree of scatter and relatively isolated locations of the LG dIrrs may provide a template for how low mass galaxies evolve without significant environmental influence.

Third, on average, dTrans appear to have formed $\gtrsim$ 45\% of their total stellar mass prior to z $\sim$ 2, and have experienced nearly constant SFRs since that time.  SFHs of the individual galaxies show modest scatter over most of their lifetimes.   dTrans with predominantly old SFHs tend to have lower present day gas fractions, while those with higher gas fraction have had more constant SFHs over most of their lifetimes.  This trend may provide insight into quenching mechanisms in low mass galaxies, which will discuss in a future paper.

Fourth, the dEs typically have an initial burst of star formation prior to $\sim$ 12 Gyr ago, followed by a nearly constant SFR.  The dEs show little scatter relative to the average value, and all exhibit declining SFHs starting at z $\sim$ 0.1 (2 Gyr ago).  The recent decline in SFRs may be related to the putative recent interaction between M31, M32, and NGC~205 \citep{choi2002}, while the decline in NGC~147 and NGC~185 may be due to their apparent paired nature \citep[e.g.,][]{vandenbergh1998, fattahi2013}.

In Panels (e) - (h), we compare the average SFHs between the different morphological types.  Here, in addition to statistical uncertainties, we also include differential systematic uncertainties as described in \S \ref{sec:sfherrors}.  To assess differences in the lifetime evolution of these systems, we examine when differences in the average SFHs exceed the relative uncertainties.  Taken at face value, this plot indicates that, on average, all morphological types formed $\sim$ 30\% of their stars prior to 12 Gyr, and are only statistically different after this epoch.  This conclusion would appear to support claims that dwarfs with present day morphologies did not share a similar progenitor, as suggested in \citet{grebel2003}.

However, the observed differences in our sample can be at least partially attributed to selection effects.  HST archival observations of dIrrs were often motivated by either recent SFH work or to collect as many stars as possible, and therefore the HST fields preferentially target the high surface brightness, central regions of the dIrrs. As a consequence, the SFHs derived from these fields are biased toward younger ages, relative to the expected gobal SFH of the galaxy i.e., the fraction of stellar mass formed at older ages is a lower limit.  In some cases, this effect is somewhat mitigated by including an additional outer field (e.g., IC~1613, Leo~A), but in others (e.g., NGC~6822, WLM), only fields that target different locations in the star-forming disk are available.  

Despite these hard to capture selection effects, we suggest that the luminous dSphs (e.g., Fornax, Leo {\sc I}) and dIrrs generally have similar, extended SFHs, while the lower luminosity dSphs (e.g., Leo {\sc IV}) do not.  Taken at face value, this suggests that a dwarf galaxy's SFHs may be a strong function of mass, but as we discuss in \S \ref{sec:masstrends}, it is likely a more complex combination of mass and environment.

In Panels (i) - (l) of Figure \ref{fig:avg_sfhs}, we over-plot simple SFH models (e.g., single old population, exponentially declining `$\tau$-models', and constant) on top of the average SFHs.  Overall, we see some correspondence between simple SFH models and the average SFHs.  The best matches are with the dSphs, whose average SFH is well-approximated by a $\tau$-model with $\tau =$ 5 Gyr. 

For other morphological types, the model matching is slightly more complex.  dIrrs, dTrans, and dEs all show a modest initial burst in their SFHs ($>$ 10-12 Gyr ago) followed by nearly constant or slightly rising SFHs.  The combination of an exponential SFH ($\tau \sim$ 3-5 Gyr) prior to 10-12 Gyr ago, followed by a constant SFH thereafter provides a reasonable approximation for average SFHs of these three groups.   This result is similar to results from our previous analysis of dwarf galaxies outside the LG \citep{weisz2011a}.

\begin{deluxetable*}{cccccccc}
\tablecaption{Star Formation Histories of each WFPC2 Field}
\tablehead{
\colhead{Galaxy} &
\colhead{Field} &
\colhead{Total Mass Formed} &
\colhead{f$_{10.1}$} &
\colhead{f$_{10.05}$} &
\colhead{f$_{10.0}$} &
\colhead{f$_{9.95}$} &
\colhead{f$_{9.9}$} \\
\colhead{} &
\colhead{} &
\colhead{(10$^6$ M$_{\odot}$)} &
\colhead{} &
\colhead{} &
\colhead{} &
\colhead{} &
\colhead{} \\
\colhead{(1)} &
\colhead{(2)} &
\colhead{(3)} &
\colhead{(4)} &
\colhead{(5)} &
\colhead{(6)} &
\colhead{(7)} &
\colhead{(8)} 
}
\startdata
Andromeda {\sc I} & u2e701 & 2.79 $^{+ 1.4 , 2.36 }_{- 1.05 , 2.79 }$ & 0.19 $^{+0.0,0.3}_{-0.03,0.19}$ & 0.34 $^{+0.04,0.52}_{-0.0,0.09}$ & 0.5 $^{+0.0,0.4}_{-0.11,0.14}$ & 0.5 $^{+0.05,0.46}_{-0.0,0.0}$ & 0.53 $^{+0.15,0.46}_{-0.0,0.0}$ \\ 
Andromeda {\sc II} & u41r01  & 1.94 $^{+ 0.83 , 1.55 }_{- 0.67 , 1.66 }$ & 0.27 $^{+0.01,0.26}_{-0.03,0.14}$ & 0.46 $^{+0.01,0.28}_{-0.02,0.1}$ & 0.48 $^{+0.05,0.32}_{-0.0,0.08}$ & 0.67 $^{+0.0,0.17}_{-0.12,0.22}$ & 0.67 $^{+0.0,0.17}_{-0.01,0.18}$ \\ 
Andromeda {\sc III} & u56901  & 1.54 $^{+ 0.48 , 1.16 }_{- 0.4 , 1.54 }$ & 0.11 $^{+0.02,0.56}_{-0.0,0.11}$ & 0.24 $^{+0.06,0.64}_{-0.0,0.0}$ & 0.67 $^{+0.01,0.23}_{-0.02,0.23}$ & 0.67 $^{+0.06,0.24}_{-0.01,0.2}$ & 0.86 $^{+0.0,0.04}_{-0.09,0.3}$ \\ 
Andromeda {\sc V} & u5c701  &  1.42 $^{+ 0.39 , 1.29 }_{- 0.44 , 1.03 }$ & 0.31 $^{+0.08,0.65}_{-0.0,0.01}$ & 0.42 $^{+0.09,0.54}_{-0.0,0.06}$ & 0.7 $^{+0.02,0.26}_{-0.05,0.33}$ & 0.76 $^{+0.08,0.21}_{-0.0,0.17}$ & 0.83 $^{+0.05,0.13}_{-0.0,0.14}$ \\ 
Andromeda {\sc VI} & u5c703  &  3.62 $^{+ 1.85 , 3.96 }_{- 1.59 , 3.08 }$ &  0.22 $^{+0.02,0.63}_{-0.0,0.03}$ & 0.47 $^{+0.0,0.43}_{-0.07,0.09}$ & 0.54 $^{+0.08,0.36}_{-0.03,0.09}$ & 0.59 $^{+0.12,0.33}_{-0.0,0.09}$ & 0.83 $^{+0.0,0.06}_{-0.09,0.29}$ \\ 
Andromeda {\sc VII} & u5c356  &  12.8 $^{+ 1.45 , 2.35 }_{- 1.95 , 5.46 }$ & 0.98 $^{+0.0,0.02}_{-0.08,0.15}$ & 0.98 $^{+0.0,0.02}_{-0.06,0.12}$ & 0.98 $^{+0.0,0.02}_{-0.05,0.06}$ & 0.98 $^{+0.0,0.02}_{-0.05,0.05}$ & 0.98 $^{+0.0,0.02}_{-0.04,0.04}$ \\ 
Andromeda {\sc XI} & u9x701  &  0.17 $^{+ 0.06 , 0.07 }_{- 0.05 , 0.12 }$ & 0.78 $^{+0.0,0.03}_{-0.13,0.43}$ & 0.78 $^{+0.0,0.06}_{-0.1,0.45}$ & 0.78 $^{+0.03,0.17}_{-0.07,0.3}$ & 0.78 $^{+0.03,0.17}_{-0.05,0.14}$ & 0.78 $^{+0.04,0.17}_{-0.0,0.07}$ \\ 
Andromeda {\sc XII} & u9x706  & 0.06 $^{+ 0.06 , 0.07 }_{- 0.04 , 0.06 }$ & 0.4 $^{+0.08,0.1}_{-0.07,0.27}$ & 0.4 $^{+0.1,0.14}_{-0.03,0.29}$ & 0.4 $^{+0.12,0.31}_{-0.01,0.21}$ & 0.4 $^{+0.09,0.37}_{-0.06,0.13}$ & 0.4 $^{+0.29,0.48}_{-0.0,0.12}$ \\ 
Andromeda {\sc XIII} & u9x712  &  0.16 $^{+ 0.05 , 0.05 }_{- 0.04 , 0.12 }$ & 0.9 $^{+0.0,0.0}_{-0.16,0.6}$ & 0.9 $^{+0.0,0.0}_{-0.07,0.57}$ & 0.9 $^{+0.0,0.04}_{-0.08,0.37}$ & 0.9 $^{+0.0,0.04}_{-0.05,0.31}$ & 0.9 $^{+0.0,0.05}_{-0.05,0.29}$ \\ 
Carina & u2lb01  &  0.03 $^{+ 0.02 , 0.03 }_{- 0.02 , 0.03 }$ & 0.26 $^{+0.06,0.06}_{-0.12,0.26}$ & 0.26 $^{+0.08,0.12}_{-0.05,0.18}$ & 0.46 $^{+0.0,0.0}_{-0.07,0.26}$ & 0.46 $^{+0.0,0.07}_{-0.07,0.22}$ & 0.46 $^{+0.0,0.07}_{-0.06,0.22}$ 
\enddata
\tablecomments{The cumulative SFHs for each field in our sample. Column (3) is the total stellar mass formed per field computed by integrating the SFH.  Note that this quantity is not the present day stellar mass, as it does not account for stellar evolutions effects (e.g., mass loss, stellar death).  Columns (4) - (8) list the faction of total stellar mass formed prior to the specified epoch from the best fit SFH.  The upper and lower error bars reflect the 16th and 84th percentiles in the uncertainties, with the random component listed first, and the total uncertainty (random plus systematic) listed second. By construction, the cumulative SFHs are 0 at $\log(\mathrm{t}) =$ 10.15 and 1 at $\log(\mathrm{t}) =$ 6.6.  Note that this table is only a sample of the galaxies and time bins.  The full table is available in electronic format on this website:  \textcolor{blue}{\url{http://people.ucsc.edu/~drweisz}}.} 
\label{tab:field_sfhs}
\end{deluxetable*}

\begin{deluxetable*}{cccccccc}
\tablecaption{Total Star Formation Histories of Galaxies with Multiple Fields}
\tablehead{
\colhead{Galaxy} &
\colhead{Total Mass Formed} &
\colhead{f$_{10.1}$} &
\colhead{f$_{10.05}$} &
\colhead{f$_{10.0}$} &
\colhead{f$_{9.95}$} &
\colhead{f$_{9.9}$} \\
\colhead{} &
\colhead{(M$_{\odot}$)} &
\colhead{} &
\colhead{} &
\colhead{} &
\colhead{} &
\colhead{} \\
\colhead{(1)} &
\colhead{(2)} &
\colhead{(3)} &
\colhead{(4)} &
\colhead{(5)} &
\colhead{(6)} &
\colhead{(7)} 
}
\startdata
IC~1613 & 5.62 $^{+ 1.97 , 4.22 }_{- 1.96 , 5.32 }$ & 0.14 $^{+0.01,0.16}_{-0.02,0.08}$ & 0.25 $^{+0.02,0.1}_{-0.0,0.11}$ & 0.3 $^{+0.01,0.1}_{-0.02,0.13}$ & 0.31 $^{+0.02,0.24}_{-0.01,0.11}$ & 0.31 $^{+0.03,0.23}_{-0.01,0.07}$ \\ 
Leo~A & 2.19 $^{+ 1.06 , 1.49 }_{- 1.05 , 1.97 }$ & 0.23 $^{+0.01,0.07}_{-0.07,0.17}$ & 0.3 $^{+0.0,0.03}_{-0.08,0.19}$ & 0.3 $^{+0.0,0.01}_{-0.05,0.2}$ & 0.3 $^{+0.01,0.06}_{-0.04,0.16}$ & 0.31 $^{+0.01,0.01}_{-0.03,0.24}$ \\ 
NGC~185 & 61.6 $^{+ 14.57 , 37.06 }_{- 14.93 , 53.93 }$ &  0.29 $^{+0.0,0.27}_{-0.01,0.07}$ & 0.29 $^{+0.01,0.27}_{-0.01,0.09}$ & 0.3 $^{+0.02,0.37}_{-0.0,0.06}$ & 0.68 $^{+0.01,0.3}_{-0.05,0.16}$ & 0.7 $^{+0.02,0.24}_{-0.01,0.14}$ \\ 
NGC~205 & 84.13 $^{+ 37.66 , 67.86 }_{- 30.59 , 73.11 }$ &  0.34 $^{+0.0,0.22}_{-0.04,0.1}$ & 0.34 $^{+0.03,0.33}_{-0.03,0.05}$ & 0.34 $^{+0.06,0.34}_{-0.01,0.05}$ & 0.43 $^{+0.1,0.31}_{-0.0,0.0}$ & 0.72 $^{+0.0,0.05}_{-0.12,0.3}$ \\ 
NGC~6822 & 17.54 $^{+ 5.57 , 14.32 }_{- 5.76 , 14.18 }$ & 0.19 $^{+0.01,0.13}_{-0.01,0.06}$ & 0.2 $^{+0.01,0.17}_{-0.01,0.03}$ & 0.23 $^{+0.01,0.09}_{-0.01,0.1}$ & 0.24 $^{+0.01,0.17}_{-0.01,0.08}$ & 0.24 $^{+0.02,0.22}_{-0.0,0.07}$ \\ 
Phoenix & 1.88 $^{+ 0.63 , 0.88 }_{- 0.67 , 1.37 }$ & 0.35 $^{+0.01,0.09}_{-0.02,0.07}$ & 0.45 $^{+0.02,0.06}_{-0.02,0.14}$ & 0.55 $^{+0.03,0.05}_{-0.02,0.11}$ & 0.64 $^{+0.0,0.01}_{-0.03,0.14}$ & 0.64 $^{+0.03,0.05}_{-0.0,0.08}$ \\ 
Sagittarius & 0.03 $^{+ 0.01 , 0.02 }_{- 0.01 , 0.03 }$ &  0.14 $^{+0.0,0.0}_{-0.07,0.14}$ & 0.14 $^{+0.01,0.11}_{-0.03,0.12}$ & 0.14 $^{+0.04,0.18}_{-0.0,0.1}$ & 0.14 $^{+0.04,0.21}_{-0.0,0.09}$ & 0.14 $^{+0.06,0.26}_{-0.0,0.09}$ \\ 
Sex~A & 22.42 $^{+ 9.44 , 15.69 }_{- 9.06 , 20.75 }$ & 0.61 $^{+0.0,0.18}_{-0.24,0.43}$ & 0.61 $^{+0.0,0.18}_{-0.11,0.14}$ & 0.61 $^{+0.01,0.17}_{-0.06,0.09}$ & 0.61 $^{+0.02,0.17}_{-0.04,0.08}$ & 0.61 $^{+0.04,0.17}_{-0.02,0.07}$ \\ 
Sculptor & 0.01 $^{+ 0.002 , 0.002 }_{- 0.001 , 0.002 }$ & 0.2 $^{+0.05,0.6}_{-0.06,0.06}$ & 0.84 $^{+0.05,0.16}_{-0.01,0.05}$ & 0.96 $^{+0.0,0.03}_{-0.03,0.17}$ & 0.96 $^{+0.0,0.03}_{-0.02,0.17}$ & 0.96 $^{+0.0,0.03}_{-0.01,0.12}$ \\ 
WLM & 8.63 $^{+ 2.54 , 4.49 }_{- 2.38 , 6.73 }$ & 0.32 $^{+0.0,0.08}_{-0.05,0.11}$ & 0.32 $^{+0.0,0.07}_{-0.04,0.1}$ & 0.33 $^{+0.0,0.07}_{-0.03,0.1}$ & 0.36 $^{+0.0,0.08}_{-0.04,0.08}$ & 0.36 $^{+0.0,0.08}_{-0.03,0.08}$ 
\enddata
\tablecomments{The same as Table \ref{tab:field_sfhs}, only for the combined SFHs of galaxies with multiple fields.  The total SFHs per galaxy are the unweighted sum of the SFHs of their constituent individual fields. Note that this table is only a sample of the galaxies and time bins.  The full table is available in electronic format on this website:  \textcolor{blue}{\url{http://people.ucsc.edu/~drweisz}}.} 
\label{tab:whole_sfhs}
\end{deluxetable*}

\subsection{Trends in Star Formation History as a Function of Galaxy Stellar Mass}
\label{sec:masstrends}

\begin{figure}
\begin{center}
\plotone{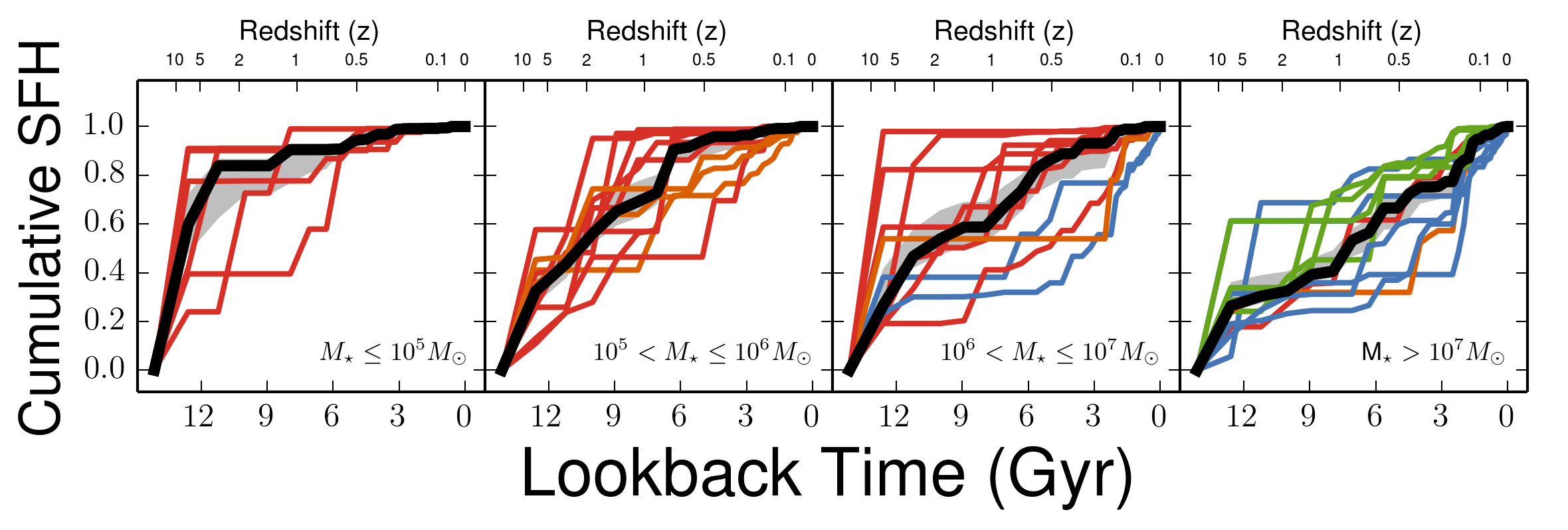}
\caption{The cumulative SFHs as a function of stellar mass.  Individual galaxy SFHs are color-coded according to morphological type (dSph, red; dIrr, blue; dTrans, orange; dE, green).  The black line is the median SFH in each mass bin and the grey uncertainty envelope is the root square sum of the differential systematic and the standard error in the median.  There is a clear trend in the SFHs such that higher mass galaxies form a larger fraction of their mass at later times.  However, this `upsizing' trend can largely be attributed to environmental effects. See \S \ref{sec:masstrends} for further discussion.}
\label{fig:mass_avg_sfhs}
\end{center}
\end{figure}

We now consider differences in our SFHs as a function of galaxy stellar mass.  For simplicity, we place the galaxies into 4 bins of stellar mass from M$_{\star}$ $\sim$ 10$^4$ - $<$ 10$^8$ \msun, and plot their SFHs in Figure \ref{fig:mass_avg_sfhs}.  The median best fit SFHs for each mass range are shown in black and the individual SFHs are color-coded by morphological type.

This plot shows a clear trend in average SFH, such that the galaxies with M$_{\star} < 10^5$ \msun\ formed $\sim$ 80\% of their stellar mass prior to z $\sim$ 2, while those with M$_{\star} > 10^7$ \msun\ formed only $\sim$ 30\% by the same epoch.  Naively, this trend could be interpreted as evidence against an extension of galaxy `downsizing' \citep[e.g.,][]{cowie1996, delucia2004, thomas2005} to the lowest masses.  If the galaxy downsizing trend persisted to low masses, we would expect to see the highest mass galaxies stop star formation before the lower mass galaxies \citep[e.g.,][]{fontanot2009}.  Here, however, the trend is reversed such it is that the lowest mass galaxies have truncated their star formation at earlier times.

However, the environment of the LG complicates this simple interpretation.  The lowest mass galaxies in this plot are all located within the virial radius of MW or of M31 and were likely subject to environmental processing (e.g., ram pressure stripping, tidal interactions) that quenched star formation at early times.  In contrast, more massive dwarfs are located well outside the virial radii of the MW or M31, and are less likely to have been significantly influenced by environment.  This type of effect has also been observed in comparisons of satellite and field galaxy properties outside the LG \citep[e.g.,][]{wetzel2013}. 

\begin{figure}
\begin{center}
\plotone{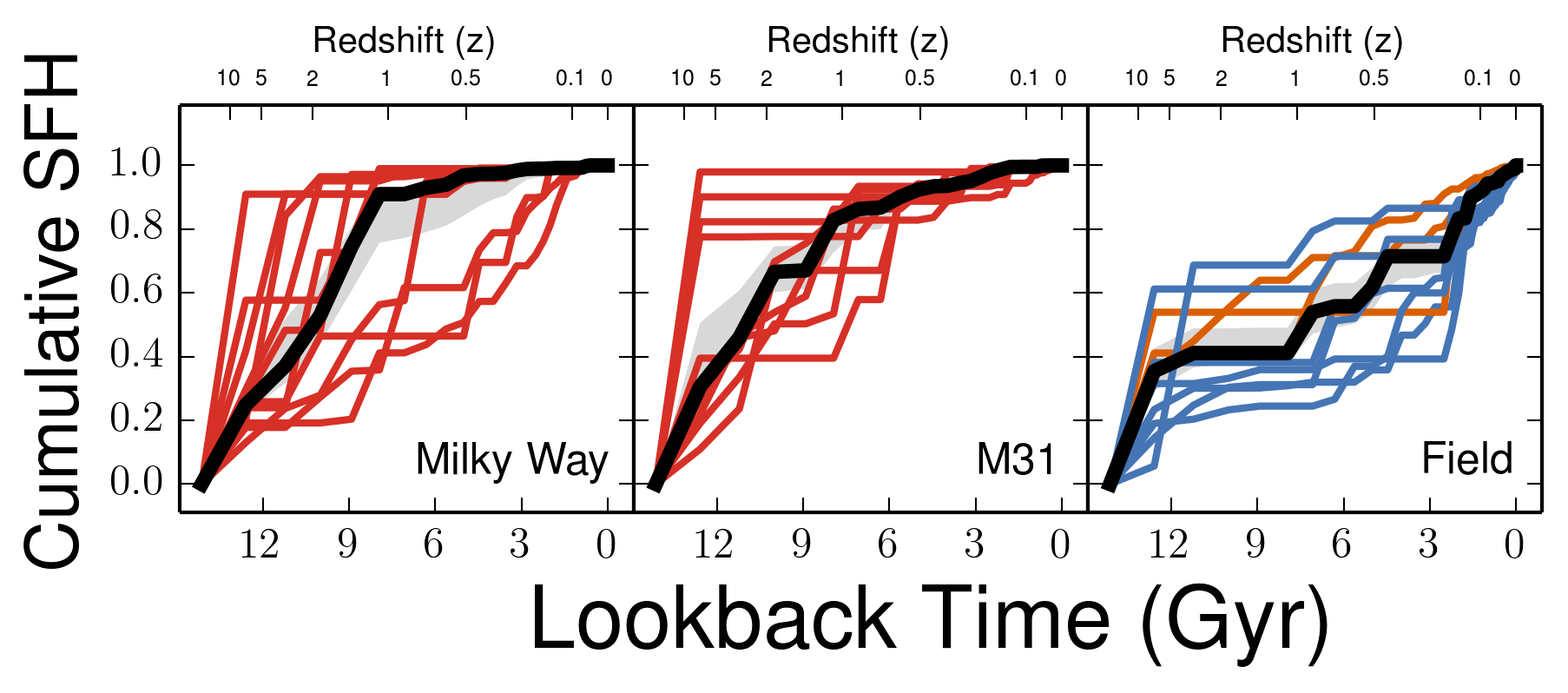}
\caption{Same as Figure \ref{fig:mass_avg_sfhs} only for the MW dSphs, M31 dSphs, and `field' galaxies. There may be subtle differences in the SFHs of MW and M31 satellites.  However, both selection effects and large uncertainties on the M31 satellites SFHs compromise any quantitative comparisons.}
\label{fig:mwm31_avg_sfhs}
\end{center}
\end{figure}

We consider environmental influences in more detail in Figure \ref{fig:mwm31_avg_sfhs}.  Here we show the best fit SFHs for dSphs located within the virial radius of the MW or M31, along with star-forming dwarfs located in the `field' of the LG. Comparing the average SFHs among the three groups, we see that all formed similar fractions of stellar mass prior to $\sim$ 10-12 Gyr ago.  Subsequently, we see that the satellite galaxies have roughly constant SFHs between $\sim$ 10-12 and 6-7 Gyr ago, after which star formation appears to drastically diminish.  In contrast, the field galaxies, on average, show little mass growth from $\sim$ 12 to 7 Gyr ago, but exhibit rising SFHs from 7 Gyr ago to the present.  As previously discussed, some of these differences could be due to field placement biases in the dIrrs, such that we are preferentially measuring a SFH biased toward younger ages.  Correcting for this effect would likely bring the average field and satellite galaxies into agreement for a longer interval at early times.  However, it is unlikely to drastically change the SFHs within the last few Gyr where there is already clear divergence among the average SFHs. 

Beyond the average SFHs, the distribution of individual SFHs in the three panels is quite interesting.   For example, MW satellites show large scatter in their SFHs, whereas the scatter in the M31 satellites and field galaxy SFHs is noticeably less.  From a physical standpoint, this scatter may reflect different in the mass assembly histories between the different sub-groups and the field population \citep[as was previously suggested based on differences in HB morphologies, e.g.,][]{dacosta2002}, or due to different manifestations of external processes \citep[e.g., heterogenous effects of reionization,][]{busha2010}. True differences between the histories of the sub-groups would be particularly interesting, as many of our cosmological models of galaxy formation are based on the properties and histories of MW satellites.  

Our data indicates that differences in the two groups may be subtle.  As shown in Figure \ref{fig:mwm31_avg_sfhs}, many of the M31 dwarfs in our sample appear to have enhanced intermediate age star formation relative to the MW satellites.  However,  a combination of shallow M31 satellite CMDs and complicated selection effects preclude us from drawing any statistically significant conclusions.  Similarly for the field galaxies, the lack of predominantly old galaxies is intriguing and may provide guidance for understanding SFHs of isolated low mass galaxies.  However, only two dIrrs have CMDs that include the oldest MSTO, meaning that our understanding of star formation in these galaxies at early times is very uncertain. 

These issues can only be addressed with deeper imaging of dwarfs outside the virial radius of the MW.  Unfortunately, the number of LG dwarfs whose SFHs are derived from the oldest MSTO is small, and as a result, our knowledge of the earliest epochs of star formation outside of MW satellites is limited.  Currently, the only galaxies with CMDs that include the oldest MSTO outside the MW virial radius are Cetus, Tucana, Leo~A, LGS~3, IC~1613, Phoenix, and Leo~T, roughly 10\% of the known LG dwarfs population.  Even less is known about the the dSphs in the M31 system, where only recently has HST imaging captured the first oldest MSTOs in two M31 satellites, Andromeda {\sc II} and Andromeda {\sc XVI} \citep{weisz2014m31}.

Finally, we examine the SFHs of dSphs.  Many dSphs are known to host complex and extended histories, and the relationship between these SFHs and their evolution is still not well-understood \citep[e.g.,][and references therein]{mateo1998, tolstoy2009}.  In contrast, the lowest luminosity dSphs are thought to generally host simple stellar populations, and are among the best candidates for remnant primordial galaxies \citep[e.g.,][]{simon2007, kirby2008, penarrubia2008, frebel2010a, belokurov2013, walker2013}.  Despite their intriguing nature, few quantitative comparisons of SFHs across the entire LG dSph spectrum exist. 

\begin{figure}
\begin{center}
\plotone{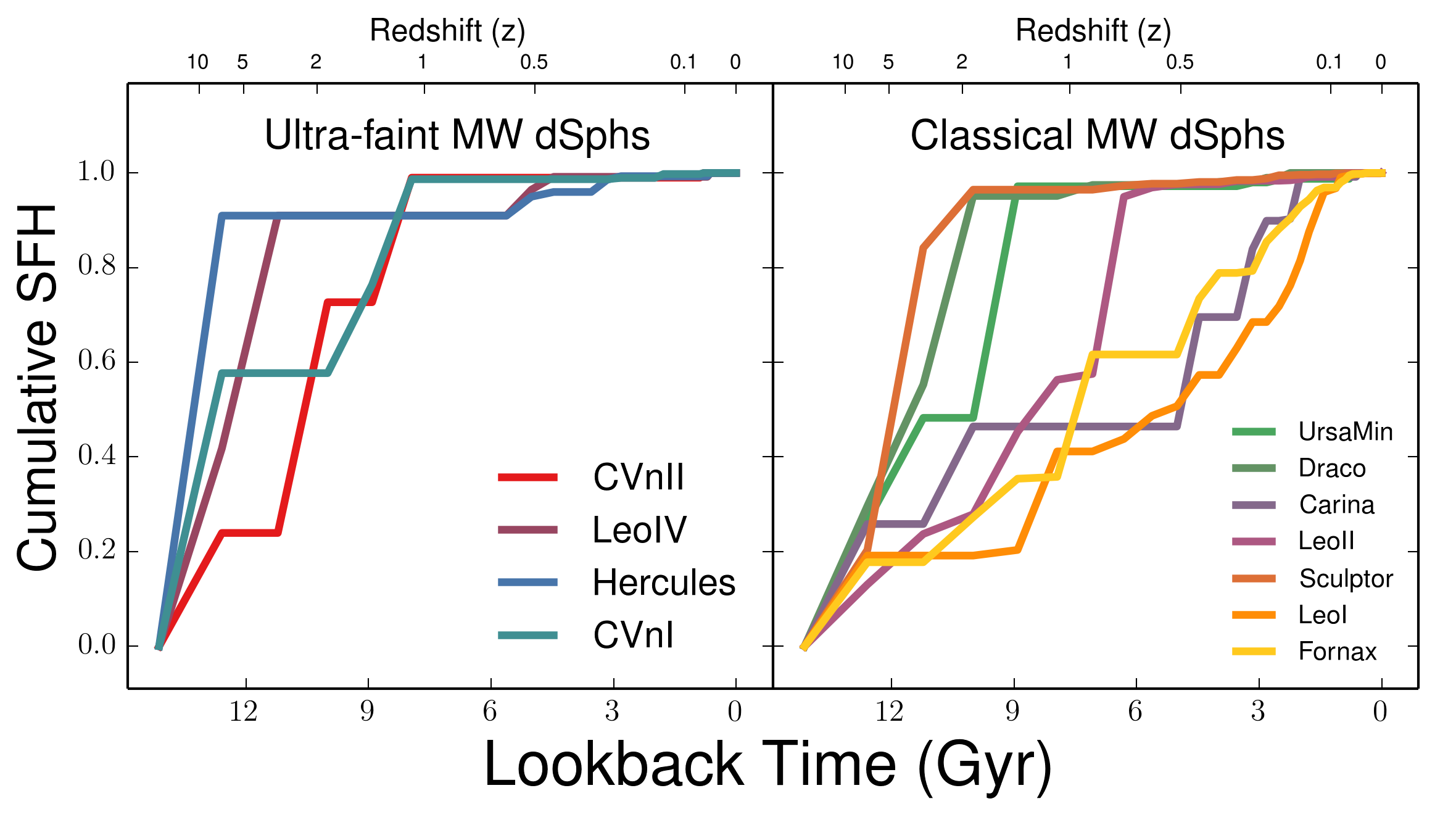}
\caption{The cumulative SFHs of the MW ultra-faint and classical dSphs.  The classical dwarfs show a diversity in SFHs: Hercules and Leo {\sc IV} are purely old, while CVn {\sc I} and CVn {\sc II} have extended SFHs.  The classical dwarfs also have both predominantly old galaxies (e.g., Sculptor, Draco) and those with extended SFHs (e.g., Carina, Fornax).  There are no apparent systematic differences between the SFHs of the ultra-faint and classical dwarfs in our sample.}
\label{fig:faint_sfhs}
\end{center}
\end{figure}

To facilitate a comparison among the dSphs, in Figure \ref{fig:faint_sfhs}, we plot the best fit SFHs of the MW dSphs.  We include only the MW dSphs as all their CMDs extend below the oldest MSTO, providing secure lifetime SFHs with minimal systematic uncertainties.  We have also divided the dSph SFHs into two groups, so-called `classical' and `ultra-faint' dwarfs, to allow for additional comparison between these two purported `classes'.  We have excluded Sagittarius from this comparison due to its unusual interaction history.

We first consider broad trends as a function of luminosity.  In general, the least luminous dSphs tend to have had their star formation quenched at earlier times than the most luminous dSphs.  For example, galaxies such as Hercules (M$_{\mathrm{V}} = -$6.6) and Leo {\sc IV} (M$_{\mathrm{V}} = -$5.8) formed $>$ 90\% of their stellar mass prior to $\sim$ 11-12 Gyr ago, while more luminous dSphs such as Fornax (M$_{\mathrm{V}} = -$13.4) and Leo {\sc I} (M$_{\mathrm{V}} = -$12.0) did not form the same percentage of their stellar mass until $\lesssim$ 3 Gyr.  These differences are likely due to variations in infall times and environmental processing effects on MW satellite galaxies \citep[e.g.,][]{lokas2012, rocha2012, kazantzidis2013}.  However, others have suggested that reionization can play an increasingly important role in the quenching of star formation in the lowest mass galaxies \citep[e.g.,][]{bullock2000, ricotti2005, bovill2011a, bovill2011b, brown2012, okamoto2012}.  We will discuss the topics of reionization (\citealt{weisz2014b}) and quenching in LG dwarfs in future papers in this series. 

There are clear exceptions to the general trend.  For example, low luminosity galaxies CVn {\sc I} (M$_{\mathrm{V}} = -$6.6) and CVn {\sc II} (M$_{\mathrm{V}} = -$4.9) have SFHs that extend to lookback times of $\sim$ 8-11 Gyr ago, while Sculptor (M$_{\mathrm{V}} = -$11.1) formed 90\% of its stellar prior to 10-11 Gyr ago.  Further, galaxies with comparable luminosities can have different SFHs: Ursa~Minor (M$_{\mathrm{V}} = -$8.8) and Draco (M$_{\mathrm{V}} = -$8.8) formed 90\% of their stellar mass prior to 8-9 Gyr ago, while Carina (M$_{\mathrm{V}} = -$9.1) formed the same amount of stellar mass $\sim$ 3-4 Gyr ago.  

It is clear from this comparison that more luminous dwarfs generally have extended SFHs. However, deviations from this relationship (e.g., CVn {\sc I}, Carina, Sculptor) suggest that the environmental history of individual galaxies are also important \citep[e.g., tidal and ram pressure stripping,][]{mayer2001, mayer2006}, and should be considered in any evolutionary model of dSphs.

Finally, we briefly comment on the dichotomy between `ultra-faint' and `classical' dSphs.  The unfortunate naming convention implies the presence of a systematic physical division between the two `classes'. However, most observational evidence indicates that both sample sequentially fainter regions of the dSphs luminosity function, e.g., they form a continuum on the mass-metallicity and size-luminosity relationship \citep[e.g.,][and references therein]{mcconnachie2012, belokurov2013, walker2013}, suggesting that the main differences are simply in the naming convention.

In Figure \ref{fig:faint_sfhs}, we see that there is not always a clear difference between the SFHs of ultra-faint and classical dSphs.  As previously discussed, ultra-faint galaxies such as CVn~{\sc I} and CVn~{\sc II} have SFHs that extend to lookback times of $\sim$ 8-11 Gyr ago, which is similar to several classical dwarfs, and in agreement with the conclusions of \citet{dejong2008a}.

Based on the aggregate SFH evidence from this paper and other studies of classical dwarfs \citep[e.g.,][]{dejong2008a, sand2009, sand2010, sand2012, brown2012, okamoto2012}, we suggest that a division between ultra-faint and classical dSphs is arbitrary. While the least luminous dSphs may be more susceptible to galaxy evolutionary mechanisms \citep[e.g., tidal effects, stellar feedback,][]{belokurov2009, kirby2013b} and have had simpler chemical enrichment histories \citep[e.g.,][]{frebel2012}, these effects may be a function of mass and environment, as opposed to a clean separation as implied by the naming convention.  We therefore suggest that the current dividing line between ultra-faint and classical dSphs is arbitrary.

\section{Connecting Low-z and High-z Star Formation Histories}
\label{sec:sdss}

\subsection{A Comparison with Leitner 2012}
\label{sec:leitner}

\citet{leitner2012} have recently highlighted an interesting tension between the CMD-based SFHs of nearby low mass galaxies from \citet{weisz2011a, weisz2011b} and the spectra-based SFHs of more massive star-forming SDSS galaxies from \citet{tojeiro2007,tojeiro2009}.  Despite having the lowest average masses ($\log(\langle M_{\star}/ M_{\odot} \rangle) =$ 7.6), the star-forming dwarfs from the ACS Nearby Galaxy Survey Treasury program \citep[ANGST;][]{dalcanton2009, weisz2011a, weisz2011b} have SFHs that most closely resembled the highest mass galaxies in the SDSS sample ($\log(\langle M_{\star}/ M_{\odot} \rangle) =$ 11.3).  Specifically, the  dwarfs in the $\lesssim$ 4 Mpc local volume formed $>$ 50\% of their stellar mass prior to z $\sim$ 1 whereas SDSS galaxies between $\log(\langle M_{\star}/ M_{\odot} \rangle) =$ 8.3 and 10.8 did not form comparable mass fractions until z $\sim$ 0.3 and 0.5.  Thus, the nearby dwarfs appear inconsistent with the clear `downsizing' trend that is present in the SDSS galaxy sample. \citet{leitner2012} discuss possible physical and/or technical reasons for the discrepancy (e.g., environmental influence on local galaxies, different biases in both CMD and spectral fitting), but find no clear resolution.

One weakness of the ANGST sample is the uncertainty in the SFHs for redshifts higher than z $\sim$ 1.  The typical ANGST CMD only partially resolves the RC, which results in uncertainties of order $\sim \pm$ 0.3-0.4 in the cumulative SFHs for lookback ages $\gtrsim$ 7.6 Gyr ago \citep[z $\sim$ 1;][]{weisz2011a, weisz2011b, dolphin2012}.  Thus, as suggested by \citet{leitner2012}, the conclusions of that comparison should be treated with the appropriate caution.

Here, we re-visit the \citet{leitner2012} comparison using our better constrained LG dIrr SFHs.  These galaxies are similar to the ANGST sample in the average stellar mass and they are are primarily drawn from an `isolated' intra-group requirement; many of the ANGST dIrrs are from the M81 group.  However, the LG dIrrs have CMDs that can reach 1-2 mags deeper, leading to better constraints on the SFHs for z$>$1.  However, this gain is somewhat offset by the smaller sample size available in the LG (N$_{\mathrm{LG}}$ $=$ 8, N$_{\mathrm{ANGST}}$ $=$ 25) and the fact that the average covering faction of the LG galaxies is 5-10 times less than it is for the ANGST galaxies, meaning the SFHs are less globally representative (e.g., the LG observations do not typically include halo populations).  We will discuss each of these effects in turn.

\begin{figure}
\begin{center}
\plotone{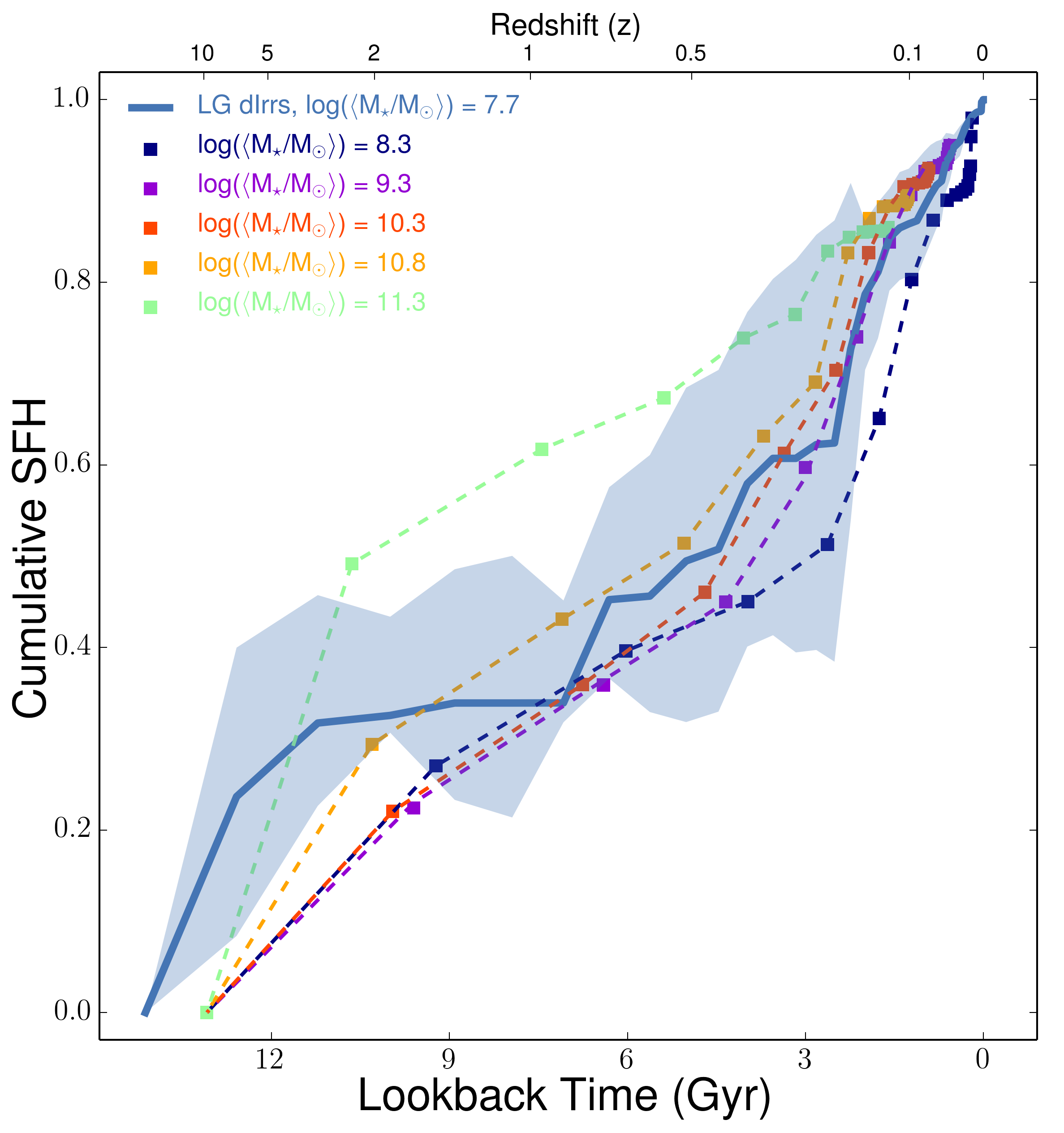}
\caption{SFHs of the LG dIrrs (solid blue line and blue error envelope) and those of star-forming SDSS galaxies (dashed) presented in \citet{leitner2012}. For the LG dIrrs, we have plotted the median dIrr SFH and the approximate standard error in the median as the error envelope. The LG dIrrs do not appear to follow the clear downsizing trend that is present in the SDSS galaxies.  They form a higher than expected stellar mass fraction at early times, which suggests that low mass galaxies may be more efficient at forming stars.  However, there are a number of competing effects including field placement biases and systematic uncertainties that influence but do no resolve this tension.  We discuss this comparison in detail in \S \ref{sec:sdss}.}
\label{fig:leitner_sfhs}
\end{center}
\end{figure}

In Figure \ref{fig:leitner_sfhs}, we show the median cumulative SFH along with statistical and systematic uncertainties for our eight LG dIrrs.  We have over-plotted the SDSS SFHs from \citet{leitner2012}, and have excluded the dTrans galaxies from the LG comparison sample.  

The trends in Figure \ref{fig:leitner_sfhs} are similar to what was found with the ANGST sample.  Namely that, on average, the LG dIrrs formed a higher fraction of their stellar mass at earlier times than most of the spectroscopically derived SFHs for the SDSS galaxies.  Specifically, on average, the LG dIrrs formed $\sim$ 30\% of their stellar mass by 10-11 Gyr ago (z $\sim$ 2).  In comparison, the most massive SDSS galaxies (log($\langle$M$_{\star}$/M$_{\odot}$$\rangle$) = 11.3) formed a comparable fraction at the same epoch.  However, for SDSS galaxies with lower stellar masses, the epoch at which they formed 30\% of their mass becomes younger. This epoch is 9-10 Gyr ago for galaxies with log($\langle$M$_{\star}$/M$_{\odot}$$\rangle$) = 10.8, and $\sim$ 7 Gyr ago for galaxies with log($\langle$M$_{\star}$/M$_{\odot}$$\rangle$) = 8.3 - 9.8. Subsequent to 7 Gyr ago (z $\sim$ 1), the LG dIrrs show a rising SFH similar to the lower mass SDSS galaxies. 

However, this finding is not as dramatic as it is for the ANGST sample, in which the dIrrs formed 50-60\% of their total stellar mass prior to 10 Gyr ago.  This is primarily due to aperture effects.  Observations of LG dIrrs cover a modest fraction of the optical body of the galaxy, and are usually centrally concentrated.  In contrast, observations of the more distant ANGST galaxies also include parts of the older halo populations, which increase the fraction of total stellar mass formed at older epoch.  Thus, fraction of stars formed in the LG dIrrs $>$ 10 Gyr ago should be considered a lower limit.

Taken at face value, this comparison has several interesting implications.  First, it suggests that nearby low mass galaxies do not follow the SFH downsizing trends that are evident in the star-forming SDSS sample examined in \citet{leitner2012}.  However, downsizing has not been investigated for galaxies in the mass range of LG dwarfs, and expectations are largely set by extrapolations from higher mass systems \citep[e.g.,][]{behroozi2013}. Thus, our LG dIrr SFHs suggest that downsizing trends may not monotonically continue with decreasing galaxy mass.  

Second, our results suggest that low mass galaxies may have been more efficient at forming stars at early times ($\gtrsim$ 10 Gyr ago) than is commonly thought.  This finding echoes the conclusions from our study of nearby dwarfs outside the LG \citep{weisz2011a}.  The current paradigm is that low mass galaxies are the least efficient at forming stars, which gives rise to nearly constant or rising SFHs \citep[e.g.,][and references therein]{behroozi2013b}, but our measurements suggest otherwise.  We discuss star formation efficiencies in more detail in \S \ref{sec:behroozi}.

Having considered physical reasons for the observed differences in the LG and SDSS SFHs, we now consider systematic effects.  One potential issue is the placement of the archival HST fields.  As discussed in \S \ref{sec:sample}, HST has typically targeted the central regions of LG dIrrs, which has two effects on the SFHs.  First, the central regions of dIrrs typically contain a higher percentage of younger stellar populations compared to the overall galaxy \citep[e.g.,][]{hidalgo2013}.  As a result, the SFHs would be biased toward younger ages. Correcting for this effect would result in a \emph{larger} discrepancy, and cannot resolve the current tension.
 
A second issue is that the CMDs of dIrrs in our sample do not reach the oldest MSTO.  As a result, there are increased systematic uncertainties at older and intermediate ages in these galaxies (see \S \ref{sec:lcidcompare}).  We have estimated the amplitude of these uncertainties on the ensemble dIrr population, and find that they are unable to resolve the tension with the \citet{leitner2012} SFHs.

Third, is that the assembly history of the LG may have influenced the ancient SFHs of these dIrrs.   The LG is not the lowest density environment in which dwarfs are formed \citep[e.g., voids,][]{pustilnik2002, pustilnik2011}, which may have affected their SFHs at early times.  The interaction between a gas-rich dwarf and a massive galaxies can result in several effects, including the temporary enhancement of star-formation, as is thought to be the case of the SMC and LMC \citep[e.g.,][]{besla2007, besla2012, weisz2013b} and the eventual  quenching of star-formation \citep[e.g.,][]{teyssier2012, slater2013}. However, such environmental influences at early times are not well-understand, and therefore it is challenging to assess if and by how much the group environment influenced the SFHs of LG dIrrs.

Finally, we note that there are also systematics associated with the spectra-based SFHs of the SDSS galaxies.  As discussed by \citet{leitner2012} and \citet{conroy2013}, SFHs of star-forming galaxies are challenging to measure from spectroscopy.  Luminous main sequence stars dominate most of a star-forming galaxy's optical spectrum and can therefore `hide' the signatures of older, lower mass stars.  Effectively, this would systematically bias all spectra-based SFHs of star-forming galaxies to younger ages.  As explained in \citet{conroy2013}, while there has been extensive testing of spectra based SFH determinations \citep[e.g,][]{cidfernandes2005, tojeiro2007}, these tests have primarily focused on internal diagnostics (e.g., self-consistency between star-forming and quiescent galaxies, establishing achievable time resolution as a function of signal-to-noise).  Comparisons with other techniques (e.g., broad band SED fitting, CMD analysis) and estimates of systematics in the modeling process are either sparse or unexplored.  This type of systematic bias and/or underestimation of uncertainties could also contribute to the observed discrepancy.  We refer the reader to \citet{conroy2013} for an extensive review of the strengths and weakness of spectra-based SFH determinations.

\subsection{A Comparison with the Behroozi Star Formation Models}
\label{sec:behroozi}

\citet{behroozi2013} combine abundance matching with extensive observational data to model the average SFHs as a function of halo mass across cosmic time.  They employ a novel probabilistic model that incorporates a wide range of data from high redshift SFRs \citep[e.g.,][]{bouwens2011, bouwens2012, bradley2012} to observations of recent star formation in the local universe \citep[e.g., \halpha, UV, 24 $\mu$m;][]{salim2007, kennicutt2008, moustakas2013}.  They demonstrate agreement between their models and the cosmic SFH and the SFHs of galaxies more massive than M$_{\star}$ $\sim$ 10$^9$ \msun.  However, they emphasize that due to a lack of data, their SFHs for galaxies with M$_{\star}$ $\sim$ 10$^9$ \msun\ are extrapolations from higher mass systems, and are therefore highly uncertain.  

\begin{figure}
\begin{center}
\plotone{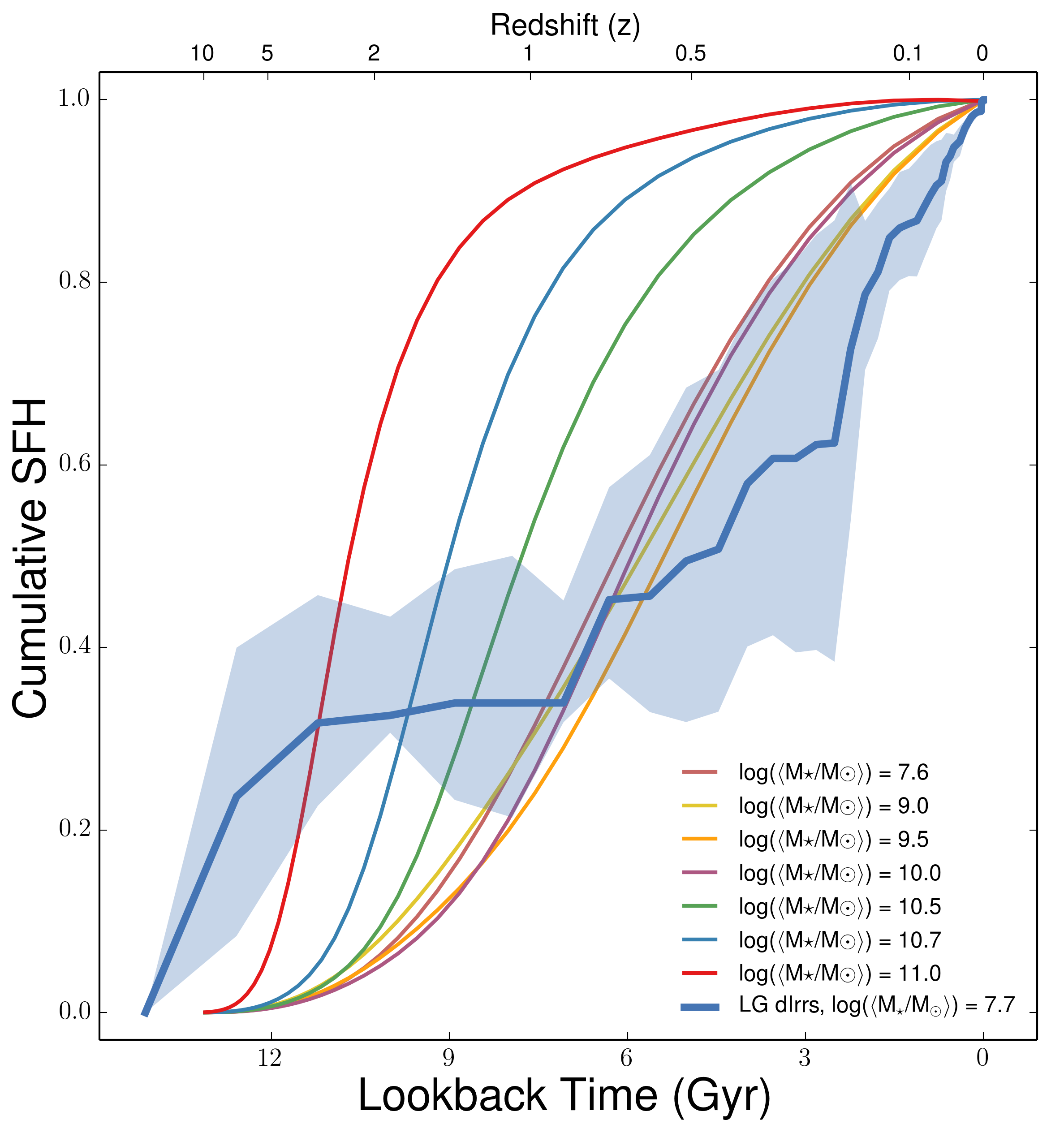}
\caption{Same as Figure \ref{fig:leitner_sfhs} only for the median SFHs per mass range from the models of \citet{behroozi2013}.  The \citet{behroozi2013} SFHs for M$_{\star}$ $\lesssim$ 10$^9$ \msun\ are extrapolations from higher mass models, where observational constraints are available.  Despite this caveat, we find reasonable agreement between the LG dIrr SFHs and the low mass SFH models for look back times younger than $\sim$ 9-10 Gyr ago.  However, at older times, the models significantly under-predict the fraction of stellar mass formed.  We discuss sources of this discrepancy in \S \ref{sec:leitner} and \S \ref{sec:behroozi}. }
\label{fig:behroozi_sfhs}
\end{center}
\end{figure}

We use our LG SFHs to explore the validity of this extrapolation at stellar masses $\lesssim$ 10$^9$ \msun.  In Figure \ref{fig:behroozi_sfhs}, we plot the median cumulative SFH of the ensemble of LG dIrrs along with statistical and systematic uncertainties.  We also over-plot the median cumulative SFHs for each stellar mass range analyzed in \citet{behroozi2013}.  When considering the full uncertainties, we find good agreement between the LG dIrr SFHs and the model predications for ages younger than $\sim$ 9 Gyr ago.  

At older times, there is significant disagreement between the observations and models.  Specifically, prior to 10 Gyr ago, the LG dIrrs formed an order of magnitude more stellar mass than is predicted by these models.  As discussed in \S \ref{sec:leitner}, correcting for biases in the measured SFHs (e.g., aperture effects) will typically \emph{increase} the amount of stellar mass formed at ancient time, which exacerbates the difference.  Similarly, including the factor of $\sim$ 2 scatter associated with the \cite{behroozi2013} and cannot reconcile the discrepancies.

From this comparison, it appears that nearby low mass galaxies have formed more stars at early times than is expected from abundance matching techniques, which are forced to rely on SFH extrapolations from higher mass galaxies. This comparison clearly demonstrates that caution is warranted when trying to infer the properties of low mass galaxies from their higher mass counter parts, as is emphasized in \citet{behroozi2013}.

As in \S \ref{sec:leitner}, the source of these discrepancies is unclear.  One physical possibility is that early star formation in low mass galaxies was more efficient than is widely believed.  At present, low mass galaxies are thought to be among the least efficient star forming systems in the universe.  For example, \citet{behroozi2013b} argue that a time-independent star formation efficiency that scales with halo mass, i.e., the least mass galaxies are also least efficient at forming stars, does well in replicating the cosmic SFH and other higher redshift observations.  But, they again emphasize a lack of data for low mass systems, and point out the dangers of extrapolating their results.

One possible explanation is that our knowledge of star formation efficiency in dwarf galaxies comes from z$=$0 populations. Most constraints on star formation efficiency for low mass galaxies come from studies of recent star formation (e.g., \halpha, UV, 24 $\mu$m) and current gas content \citep[e.g.,][]{leroy2008, salim2007, kennicutt2008, lee2011, kennicutt2012}.  Yet, star formation conditions and gas accretion rates at early times are thought to be more extreme than at z$=$0, making present day observables poor proxies for the early evolution of dwarf galaxies.  A higher efficiency of star formation at early times is qualitatively similar to the early models of dwarf galaxies by \citet{dekel1986}.  This model required high efficiencies at early times to generate sufficient stellar feedback energetics to eject cold gas, and induce morphological transformation.  However, there are few observations that can explore the possibility of a time-dependent star formation efficiency in dwarfs.  Instead, a comparison between measured SFHs and SFHs from fully cosmological simulations of dwarfs \citep[e.g.,][]{governato2010, shen2013} with different efficiencies may provide the best path forward for exploring the evolution of star formation efficiencies in low mass galaxies.  We will pursue this comparison in a future paper.

\section{Summary}
\label{summary}

We present uniformly derived SFHs of 40 LG dwarf galaxies based on analysis of CMDs constructed from archival HST/WFPC2 imaging.  We find the following results:

\begin{enumerate}
\item The lifetime SFH of a galaxy can accurately be measured from a CMD that does not reach the oldest MSTO.  SFHs from shallower data are consistent with those derived from CMD that reach below the oldest MSTO, provided that systematic uncertainties in the underlying stellar models are properly considered.  However, only CMDs that contain the oldest MSTO can be used to precisely constrain the first epoch(s) of star formation.

\item We highlight challenges in measuring the SFHs of early type galaxies due to an inadequate number of TP-AGB star evolutionary points in the Padova stellar models.  For extremely shallow CMDs, the resulting SFH can either be inaccurate or overly precise.  The majority of our sample, 38 out of 40 galaxies, are unaffected by this problem.  We show that using multiple stellar models can somewhat alleviate this problem, and note that SFHs of early type galaxies that primarily rely on TP-AGB models should be treated with caution.

\item For sparsely populated systems, e.g., ultra-faint dwarfs, increasing the survey area (or number of stars at the oldest MSTO) leads to better SFH constraints compared to a small area extremely deep CMD. SFHs from CMDs that extend well below the oldest MSTO have the same amplitude of systematic uncertainty as SFHs measured from CMDs that capture the MSTO at modest signal-to-noise, i.e., $\sim$ 1 mag below the oldest MSTO.  This is because stars below the oldest MSTO are highly insensitive to SFH.  Instead, for sparse systems, the limiting factor in SFH precision is the number of stars on the CMD. CMDs with more stars in age sensitive regions (e.g., SGB, MSTO), e.g., from surveying large areas, will lead to more precise and more globally representative SFHs compared to extremely deep CMDs of smaller areas.

\item The average dSph SFH can be approximated by an exponentially declining SFH with $\tau =$ 5 Gyr.  In contrast, the dIrrs, dTrans, and dEs have SFHs that are better fit by an exponentially declining SFH prior to $\sim$ 10 Gyr ago, with $\tau \sim$ 3-4 Gyr, and a constant or rising SFH thereafter.)

\item  The least massive LG dwarfs (M$_{\star}$ $<$ 10$^5$ \msun) formed $\gtrsim$ 80\% of their total stellar mass prior to z $\sim$ 2 (10 Gyr ago), while the more massive LG dwarfs (M$_{\star}$ $>$ 10$^5$ \msun) formed $\sim$ 30\% of their total stellar mass by the same epoch.  We discuss how this `upsizing' trend might be attributed to selection biases and environmental effects in the LG. 

\item Our data suggest subtle differences in the SFHs of M31 and MW satellites.  However, the shallow depth of the M31 dSph CMDs and complicated selection biases preclude any concrete conclusions.  Similarly, there are tantalizing differences in the early SFHs of field and satellites population, but CMDs of galaxies outside the virial radius of the MW are too shallow for the necessary constraints on the ancient SFHs.

\item  Lower luminosity dSphs are less likely to have extended SFHs than higher luminosity dSphs.  However, the modest scatter in the SFHs suggests that the trend is not a simple function of stellar mass, and that environmental effects (e.g., interaction history) have played an important role in shaping the evolutionary history of any given galaxy.

\item Some `ultra-faint' and `classical'  dSphs have similar SFHs.  We suggest that the boundary between ultra-faint and `classical' dSph is arbitrary. 

\item LG dIrrs formed a higher fraction of stellar mass prior to z=2 than the star-forming SDSS sample from \citet{leitner2012}.  We discuss possible systematics and selection effects, but find no resolution to this reverse downsizing trend.

\item Prior to z$=$2, LG dIrrs also formed a higher fraction of stellar mass than is predicted by the abundance matching models of \citet{behroozi2013b}.  This demonstrates that caution is warranted when inferring low mass galaxy properties from higher mass galaxy counterparts.  One solution to this tension is that low mass galaxies may have had higher-than-expected star formation efficiencies at early times.

\end{enumerate}

\section*{Acknowledgements}

The authors would like to thank the anonymous referee for insightful comments that helped improve the scope and clarity of this paper.  DRW would like to thank Alyson Brooks, Charlie Conroy, Sam Leitner, Nicolas Martin, Peter Behroozi, Thomas de Boer, Vasily Belokurov, Oleg Gnedin, Erik Tollerud, Andrew Wetzel, Sijing Shen, and Evan Kirby for insightful discussions about dwarf galaxy SFHs and the Local Group in general, and also Hans-Walter Rix and the MPIA for their hospitality during the assembly of this paper.  Support for DRW and KMG is provided by NASA through Hubble Fellowship grants HST-HF-51331.01 and HST-HF-51273.01, respectively, awarded by the Space Telescope Science Institute. Additional support for this work was provided by NASA through grant number HST AR-9521 from the Space Telescope Science Institute, which is operated by AURA, Inc., under NASA contract NAS5-26555. This research made extensive use of NASA's Astrophysics Data System Bibliographic Services, and also made use of NASA's Astrophysics Data System Bibliographic Services and the NASA/IPAC Extragalactic Database (NED), which is operated by the Jet Propulsion Laboratory, California Institute of Technology, under contract with the National Aeronautics and Space Administration. In large part, analysis and plots presented in this paper utilized iPython and packages from Astropy, NumPy, SciPy, and Matplotlib \citep[][]{hunter2007, oliphant2007, perez2007, astropy2013}.\\

{\it Facility:} \facility{HST (WFPC2, ACS)}

\clearpage

\appendix

\setcounter{table}{0}
\renewcommand{\thetable}{A\arabic{table}}

\LongTables
\begin{deluxetable}{lc}
\tablecolumns{2}
\tablehead{
\colhead{Reference} &
\colhead{Galaxy} \\
\colhead{(1)} &
\colhead{(2)} 
}
\tablecaption{Literature SFHs of Local Group Dwarf Galaxies}
\startdata
\citet {ferraro1989} & WLM \\
 \citet{tosi1989} & WLM, Sex~B \\
\citet{tosi1991} & Sex~B \\
\citet{greggio1993} & DDO~210 \\
\citet{smeckerhane1994} & Carina \\
 \citet{marconi1995} & NGC~6822 \\
 \citet{dacosta1996} & And~{\sc I} \\
\citet{gallart1996} & NGC~6822 \\
\citet{lee1996} & NGC~205 \\

\citet{saviane1996} & Tucana \\
 \citet{tolstoy1996} & Leo~A \\
 \citet{aparicio1997a} & LGS3 \\
 \citet{aparicio1997b} & Peg~DIG \\
 \citet{aparicio1997c} & Antlia \\
\citet{dohmpalmer1997a} & Sex~A \\
\citet{dohmpalmer1997b} & Sex~A \\

\citet {han1997} & NGC~147 \\
 \citet{mighell1997} & Carina \\
\citet{mould1997} & LGS3 \\
\citet{dohmpalmer1998} & GR~8 \\
 \citet{gallagher1998} & Peg~DIG \\
 \citet{grillmair1998} & Draco \\
\citet{hurleykeller1998} & Carina \\
\citet{martinezdelgado1998} & NGC~185 \\

\citet {tolstoy1998} & Leo~A \\
 \citet{bellazzini1999} & Sag~DIG \\
\citet{buonanno1999} & Fornax \\
\citet{caputo1999} & Leo {\sc I} \\
 \citet{cole1999} & IC~1613 \\
 \citet{gallart1999} & Leo {\sc I} \\
\citet{karachentsev1999} & Sag~DIG \\
\citet{martinezdelgado1999} & NGC~185 \\

\citet {mighell1999} & Ursa~Minor \\
 \citet{monkiewicz1999} & Sculptor \\
\citet{dacosta2000} & And {\sc II} \\
\citet{dolphin2000a} & WLM \\
 \citet{hernandez2000} & Carina, Leo {\sc I}, Leo {\sc II}, Ursa Minor\\
 \citet{holtzman2000} & Phoenix \\
\citet{saviane2000} & Fornax \\
\citet{hunter2001} & IC~10 \\

\citet {miller2001} & LGS~3 \\
 \citet{wyder2001} & NGC~6822 \\
\citet{carrera2002} & Ursa~Minor \\
\citet{dohmpalmer2002} & Sex~A \\
 \citet{dolphin2002} & Carina, Draco, Leo {\sc I}, Leo {\sc II}, Sagittarius, Sculptor, Ursa Minor\\
 \citet{sarajedini2002} & Cetus \\
\citet{schulteladbeck2002} & Leo~A \\
\citet{dacosta2002} & And {\sc III} \\

\citet {dolphin2003} & Leo~A \\
 \citet{hidalgo2003} & Phoenix \\
\citet{komiyama2003} & NGC~6822 \\
\citet{skillman2003} & IC~1613 \\
 \citet{butler2005} & NGC~185, NGC~205 \\
 \citet{momany2005} & Sag~DIG \\
\citet{bernard2007} & IC~1613 \\
\citet{cole2007} & IC~1613 \\
\citet{komiyama2007} & Leo {\sc II} \\

\citet {yuk2007} & IC~1613 \\
 \citet{dejong2007} & Canis Major \\
\citet{coleman2008} & Fornax \\
\citet{okamoto2008} & Ursa~Major {\sc I} \\
 \citet{dejong2008a} & Bootes {\sc I}, Bootes {\sc II}, CVn {\sc I}, CVn {\sc II}, Coma Ber, Hercules, \\
 & Leo {\sc IV}, Leo T, Segue 1, Ursa Major {\sc I}, Ursa Major {\sc II}, Willman 1 \\
 \citet{dejong2008b} & Leo~T \\
\citet{hidalgo2009} & Phoenix \\
\citet{lee2009} & Sextans dSph \\

\citet {monaco2009} & NGC~205\\
\citet{sand2009} & Hercules \\
\citet{sanna2009} & IC~10 \\
\citet{bono2010} & Carina \\
\citet{monelli2010b} & Cetus \\
\citet{monelli2010b} & Tucana \\
\citet{sand2010} & Leo~{\sc IV} \\
\citet{hidalgo2011} & LGS3 \\
\citet{jacobs2011} & UGC~4879 \\

\citet {pasetto2011} & Carina\\
\citet{brown2012} & Hercules, Leo {\sc IV}, Ursa Major {\sc I} \\
\citet{cannon2012} & NGC~6822 \\
\citet{clementini2012} & Leo~T \\
\citet{monachesi2012} & M32 \\
\citet{okamoto2012} & CVn {\sc I}, Bootes {\sc I}, CVn {\sc II}, and Leo {\sc IV}  \\
\citet{weisz2012b} & Leo~T \\
\citet{deboer2012a} & Sculptor \\
\citet{deboer2012b} & Fornax \\
\citet{penny2012} & Antlia \\
\citet{sand2012} & Leo {\sc V}, Pisces {\sc II}, CVn {\sc II} \\
\citet{delpino2013} & Fornax \\
\citet{mcquinn2013} & Leo~P
\enddata
\tablecomments{An exhaustive list of original CMD-based SFH studies of LG dwarfs between 1989 and 2013.  We have excluded LG SFH compilations \citep[e.g.,][]{mateo1998, tolstoy2009}, as they only provide summaries of SFHs already in the literature.}
\label{tab:full_litsfhs}
\end{deluxetable}

\end{document}